\documentclass[journal]{IEEEtran}
\IEEEoverridecommandlockouts
\usepackage{graphicx}
\usepackage{amsmath}
\usepackage{amsfonts}
\usepackage{mathrsfs}
\usepackage[OT1]{fontenc} 
\usepackage{enumerate}
\usepackage{amsthm}
\usepackage{tabularx}
\usepackage{arydshln}

\usepackage[colorlinks=true,linkcolor=blue,citecolor=blue]{hyperref}
\usepackage{mwe}
\usepackage{subfig}
\usepackage{float}
\usepackage{graphicx}
\usepackage{textcomp}
\usepackage{xcolor}
\usepackage{footnote}
\usepackage{makecell}
\usepackage{graphicx}
\usepackage{amsmath}
\usepackage{amsfonts}
\usepackage{mathrsfs}
\usepackage[OT1]{fontenc} 
\usepackage{enumerate}
\usepackage{amsthm}
\usepackage{tabularx}
\usepackage{arydshln}
\usepackage[numbers, compress]{natbib}

\usepackage{url}
\usepackage{breakurl}

\usepackage{subfig}
\usepackage{textcomp}
\usepackage{xcolor}
\usepackage{footnote}
\usepackage{makecell}
\usepackage[draft]{todonotes}

\usepackage{mathtools}
\usepackage{booktabs}
\usepackage{multirow}
\usepackage{rotating}

\usepackage{algorithm}
\usepackage{algpseudocode}
\usepackage[inline]{enumitem}

\colorlet{r1}{black}
\usepackage[numbers]{natbib}

\newcommand{\RNum}[1]{\uppercase\expandafter{\romannumeral #1\relax}}

\begin{document}
\bstctlcite{IEEEexample:BSTcontrol}
\title{Autonomous and Semi-Autonomous Intersection Management: A Survey}

\author{Zijia~Zhong,~\IEEEmembership{Member,~IEEE,}
        Mark~Nejad,~\IEEEmembership{Member,~IEEE,}
        and~Earl~E.~Lee,~\IEEEmembership{Member,~IEEE}
\thanks{Z. Zhong is with the Center for Integrated Mobility Science at the National Renewable Energy Laboratory. The work was performed when he was with the Department of Civil and Environmental Engineering, University of Delaware, e-mail: zijia.zhong@ieee.org.

M. Nejad and E. E. Lee are with the Department
of Civil and Environmental Engineering, University of Delaware, Newark,
DE, 19716 USA, e-mail: \{nejad, elee\}@udel.edu}
}

\markboth{DOI:10.1109/MITS.2020.3014074 (in press)}%
{Shell \MakeLowercase{\textit{et al.}}: Bare Demo of IEEEtran.cls for IEEE Journals}
%

\maketitle
\begin{abstract}
Intersection is a major source of traffic delays and accidents within modern transportation systems.
Compared to signalized intersection management, autonomous intersection management (AIM) coordinates the intersection crossing at an individual vehicle level, which provides additional flexibility. AIM can potentially eliminate stopping in intersection crossing due to traffic lights while maintaining a safe separation among conflicting movements. In this paper, the state-of-the-art AIM research among various disciplines (e.g., traffic engineering, control engineering) is surveyed from the perspective of three hierarchical layers: corridor coordination layer, intersection management layer, and vehicle control layer. \textcolor{r1}{The key aspects of AIM designs are discussed in details, including conflict detection schemes, priority rules, control centralization, computation complexity, etc. The potential improvements for AIM evaluation with the emphasis of realistic scenarios are provided.
This survey serves as a comprehensive review of AIM design and provides promising directions for future research.}

\end{abstract}


\section{Introduction}
Intersection is a major source of traffic delays and accidents.
According to the National Motor Vehicle Causation Survey conducted between 2005 and 2007, 36\% of the surveyed crashes (i.e., 2,188,969) were intersection related in the United States.  Among them, inadequate surveillance (44.1\%) and false assumption of other's actions (8.4\%) were the most frequent culprits \cite{national2010crash}. In 2010,
crashes at intersection in the United States amounted to US\$ 120 billion in economic costs and US\$ 371 billion in societal cost  \cite{blincoe2015economic}.

Connected and Automated Vehicles (CAVs) are expected to assume a revolutionary role in mitigating traffic accidents and congestion. CAVs encompasses connected vehicles (CVs) and automated vehicles (AVs). The former relies on the two-way wireless communication, which enables real-time information sharing and cooperation among agents within a transportation system, whereas the latter eliminates human driver errors that may potentially cause crashes, traffic flow oscillations, and shock-waves.
\textcolor{r1}{ 
Thus far, the CAV-based improvements for intersection traffic management can be categorized into two groups: 
\begin{enumerate*}[label=\roman*)]
\item the incorporation of real-time, high-resolution CAV traffic data in signalized intersections management (SIM) for enhancing the signal phase and timing (SPaT) plans and  
\item the development of signal-free autonomous intersection management (AIM) that is made possible by vehicle automation and connectivity.
\end{enumerate*}
} 
\subsection{Scope of the Survey}
Insofar, there are several notable survey papers in the literature focusing on intersection management. 
Guo et al. \cite{guo2019urban} conducted a detailed review of the integration of the CAV data into signalized intersection management. 
Guo et al.'s survey emphasized the integration of CAV technology into the existing signal control framework, for instance, vehicle platoon for signalizing. With the focus on the mixed-traffic condition and CAV-enabled augmentations, they chose to exclude signal-free intersection control, which is the focus of this paper. 
Rios-Torres and Malikopoulos \cite{rios2017survey} reviewed the intersection management and on-ramp merging from the centralized and decentralized control perspectives. Chen and Englund \cite{chen2016cooperative} reviewed studies on cooperative intersection management and highlighted the major AIM research.
Li et al. \cite{li2014survey} surveyed the traffic control system with the focus of contrasting preference, such as global planning-based versus local and self-organization-based control.

\textcolor{r1}{
A systematic review that provides the overall landscape of AIM, including design philosophies, evaluation approaches, and cross-discipline perspectives still lacks, in spite of the significant potentials of AIM. Additionally, AIM studies have been steadily emerging in the recently years owing to the rapid development of CAV technology.
There are numerous papers regarding using CAV to enhanced intersection performance. The scope of this review, however, is confined to the studies that deal with coordinating conflicting vehicle movements. Hence, intersection studies without dealing with conflicting intersection movements are excluded, such as eco-approach and departure of intersection. }

\subsection{Contribution and Organization of the Paper}
The primary contribution of the paper is the survey of the state-of-the-art research on AIM from a multi-disciplinary perspective. More specifically, this survey focuses on the following areas:
\textcolor{r1}{
\begin{itemize}
\item the transition from SIM to AIM
\item intersection conflict planning
\item vehicle control for AIM
\item AIM evaluation 
\end{itemize}}

The remainder of the paper is organized as follows: Section \ref{sec: IntxMagmt} introduces the hierarchical intersection management framework that encompasses both AIM and SIM.  Section \ref{sec: AIM} reviews the key design aspects of AIMs. The studies on the vehicle control layer are reviewed in Section \ref{sec:VehControl}, followed by the evaluation scenarios for AIM in Section \ref{sec: evalAIM}. Discussion of the future research trends are presented in Section \ref{sec: discussion}, followed by Conclusion in Section \ref{sec:conclusion}

\section{Intersection Management}
\label{sec: IntxMagmt}
\begin{figure}[h]
\begin{minipage}[h]{\columnwidth}
\centering
\subfloat[standard intersection]{\includegraphics[scale=0.4]{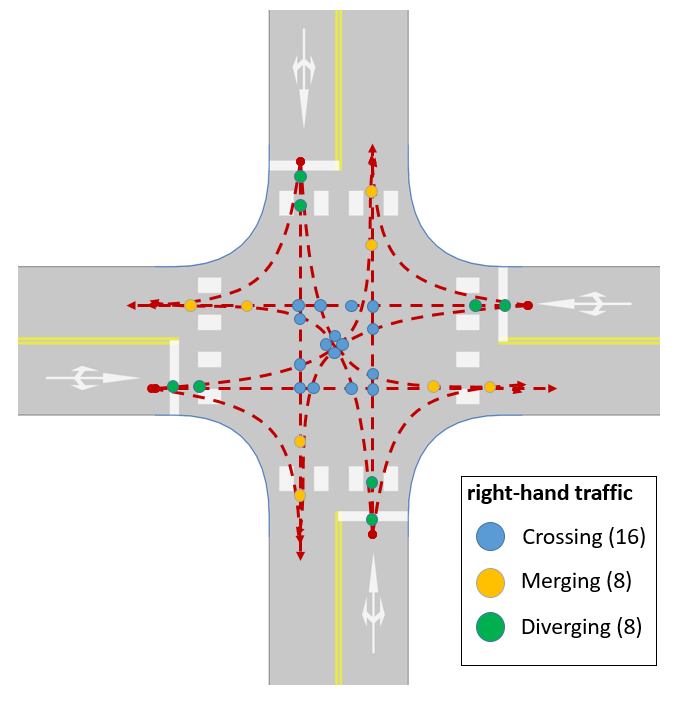}}
\end{minipage} \par
\begin{minipage}[h]{\columnwidth}
\centering
\subfloat[roundabout]{\includegraphics[scale=.4]{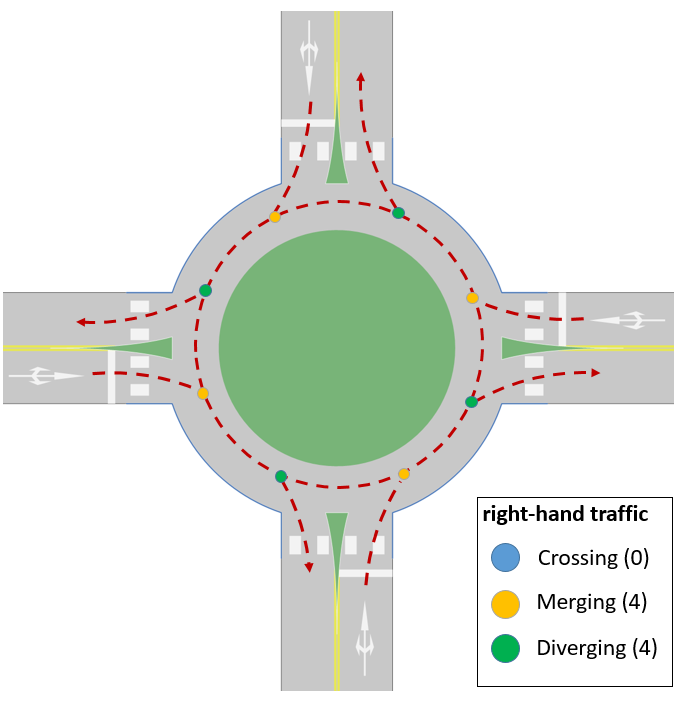}}
\end{minipage}
\caption{Intersection conflict points}
\label{fig:conflictPt}
\end{figure}

Ensuring safety by separating the time-space conflicts among approaching vehicles is of the utmost importance for intersection management. This is the case for both SIM and AIM. A conflict point is the intersection of two vehicle trajectories, where a collision could potentially occur. A standard intersection (with 4 approaches and 12 movements) has 16 conflict points as illustrated in Fig. \ref{fig:conflictPt}(a).  Roundabout is a type of non-standard intersection that are with a different set of conflict points, as shown in Fig. \ref{fig:conflictPt}(b).

\subsection{Hierarchical Layers}

\begin{figure}[h]
	\centering
	\includegraphics[width=\columnwidth]{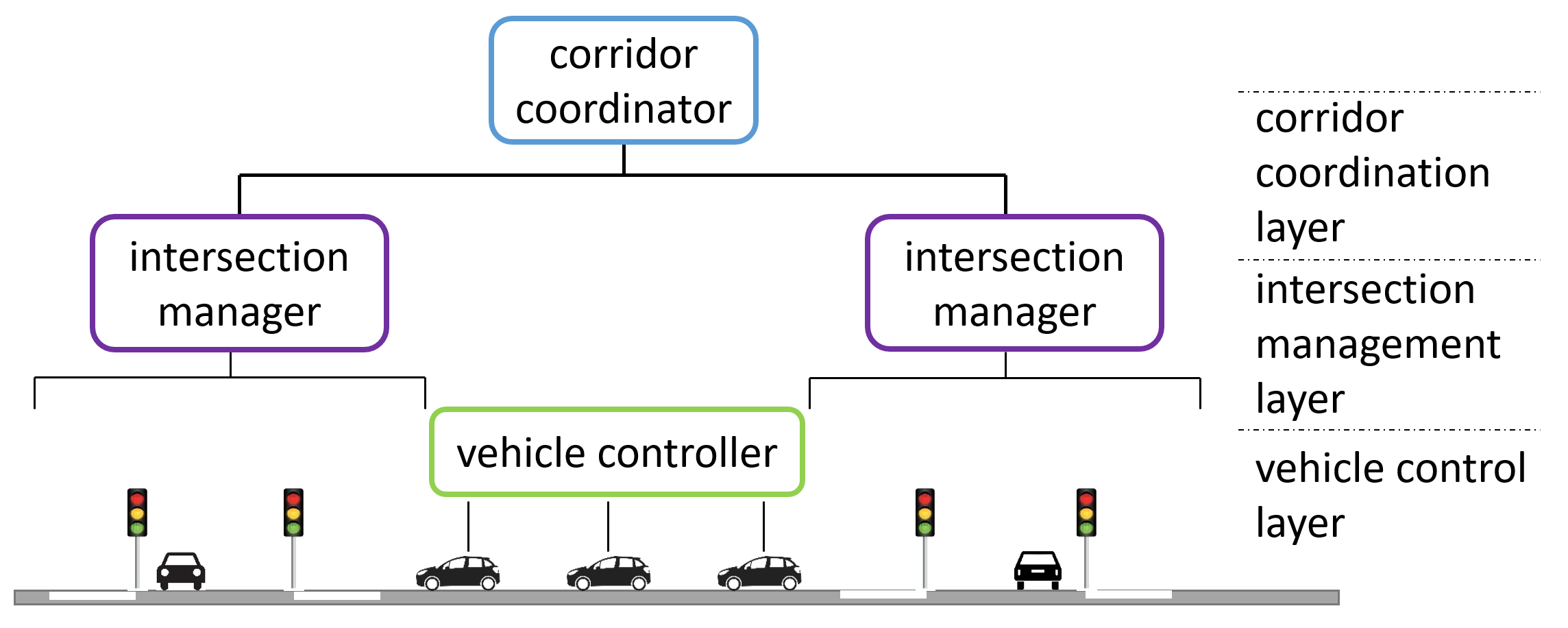}   
	\caption{Intersection control layer} 
	\label{fig: controlLayer}
\end{figure}

As illustrated in Fig. \ref{fig: controlLayer}, three hierarchical layers can be distilled from existing intersection control practices in traffic engineering, which are
\begin{enumerate*}[label=\roman*)]
\item corridor coordination layer,
\item trajectory planning layer, and
\item vehicle control layer,
\end{enumerate*}
The hierarchical framework is also suitable for evaluating AIM. The connections among the layers are enabled by communication networks. In SIM, magnetic loop detectors collect prevailing traffic conditions for the signal controllers that host the intersection management protocol (e.g., SPaT plans). The human drivers are notified with crossing permission by traffic lights. The AIM replaces the aforementioned procedure with V2X communication. 
Furthermore, the human control of vehicles is anticipated to be replaced by advanced driving assistance systems (ADAS) and eventually by automated driving systems (ADS) in AIM.

\textbf{Corridor coordination layer} deals with the coordination of multiple intersections at a corridor level. Such coordination is common for major arterials under SIM. Maximization of the green band \cite{gartner1991multi} is commonly used to ensure the progression of the major through movements across multiple intersections. Fixed-time SPaT plans among intersection are the most common approach to achieve coordination.

\textbf{Intersection management (trajectory planning) layer} assigns the crossing sequence for vehicles in AIM or vehicle groups in the case of SIM for an intersection. For SIM, the signal phases that are with conflicting movements are cycling based on a predefined phase sequence. For AIM, the intersection manager is responsible for allocating the limited time-space resources of an intersection. This aspect of AIM is often referred as trajectory planning since AIM separates conflict movements at the level of individual vehicles. 

\textbf{Vehicle Control layer} focuses on the motion control for an individual vehicle both longitudinally and laterally. Conventionally, vehicles are driven by a human who is primarily responsible for the movement of a vehicle, sometimes with the aid of vehicle sub-systems such as power steering and assisted braking. Under the CAV environment, an automated driving system is envisioned to complement and ultimately replace the driving inputs from a human.

\subsection{Roadmap for AIM}

The transition to full CAV penetration could take decades. Therefore, the three layers (illustrated in Table \ref{table:imRole}) could be assumed by various entities and levels of automation, as the development of CAV technologies progresses. The Society of Automotive Engineers International (SAE) defined six levels of vehicle automation from 0 (zero) being solely-driven by a human to 5 (five) in which the automation is in effect in any conditions \cite{SAE2002}. Thus far, with only a few exceptions, the AIM is generally not compatible with HVs. However, semi-AIMs has been proposed for the mid-level automation (Level 2 or Level 3) during the transition to full CAV penetration. The transition from SIM to AIM is accompanied by a gradual decrease in human involvement in various dynamic driving tasks.

\begin{table}[h]
\caption{Intersection Management Roadmap}
\label{table:imRole}
\resizebox{1\columnwidth}{!}
{
\begin{tabular}{l|lll}
\hline
 & \multicolumn{3}{c}{\textbf{Hierarchical Layer}} \\ \hline
\begin{tabular}[c]{@{}l@{}}Level of \\ Automation\end{tabular} & \multicolumn{1}{c}{Vehicle Control} & \multicolumn{1}{c}{\begin{tabular}[c]{@{}c@{}}Intersection \\ Management\end{tabular}} & \multicolumn{1}{c}{\begin{tabular}[c]{@{}c@{}}Corridor \\ Coordination\end{tabular}} \\ \hline
SIM (Lv. 0) & human driver & SIM & \begin{tabular}[c]{@{}l@{}}pre-timed or \\ hardwired connection\end{tabular} \\
SIM (Lv. 1) & \begin{tabular}[c]{@{}l@{}}human driver \\ + ADAS\end{tabular} & \begin{tabular}[c]{@{}l@{}}SIM +\\ Vehicle Info.\end{tabular} & \begin{tabular}[c]{@{}l@{}}pre-timed or \\ hardwired connection\end{tabular} \\
SIM (Lv.2) & \begin{tabular}[c]{@{}l@{}}human driver \\ + ADAS\end{tabular} & \begin{tabular}[c]{@{}l@{}}SIM + \\ Vehicle Info.\end{tabular} & \begin{tabular}[c]{@{}l@{}}pre-timed or \\ hardwired connection\end{tabular} \\
SIM (Lv.3) & \begin{tabular}[c]{@{}l@{}}human driver \\ + ADAS\end{tabular} & \begin{tabular}[c]{@{}l@{}}SIM or \\ Semi-AIM\end{tabular} & \begin{tabular}[c]{@{}l@{}}hardwired connection or \\ wireless communication\end{tabular} \\
Simi-AIM (Lv.4) & \begin{tabular}[c]{@{}l@{}}human driver \\ + ADAS\end{tabular} & \begin{tabular}[c]{@{}l@{}}SIM or \\ Semi-AIM\end{tabular} & \begin{tabular}[c]{@{}l@{}}hardwired connection or \\ wireless communication\end{tabular} \\
AIM (Lv.5) & ADS & AIM & wireless communication \\ \hline
\end{tabular}
}
ADAS: advanced driver-assistance systems\\
ADS: automated driving systems
\end{table}

\section{Autonomous Intersection Management}
\label{sec: AIM}
An isolated AIM is comprised of two layers: the trajectory planning layer and the vehicle control layer. To differentiate intersection management with SIM, we use the term ``trajectory planning'' exclusively for AIM in this paper.  Priority assignment and a reservation system are the two key aspects of trajectory planning, which are discussed in this section. The vehicle control layer is discussed in Section \ref{sec:VehControl}.

\subsection{Time-space Reservation}
\begin{figure}[h]
	\centering
	\includegraphics[width=\columnwidth]{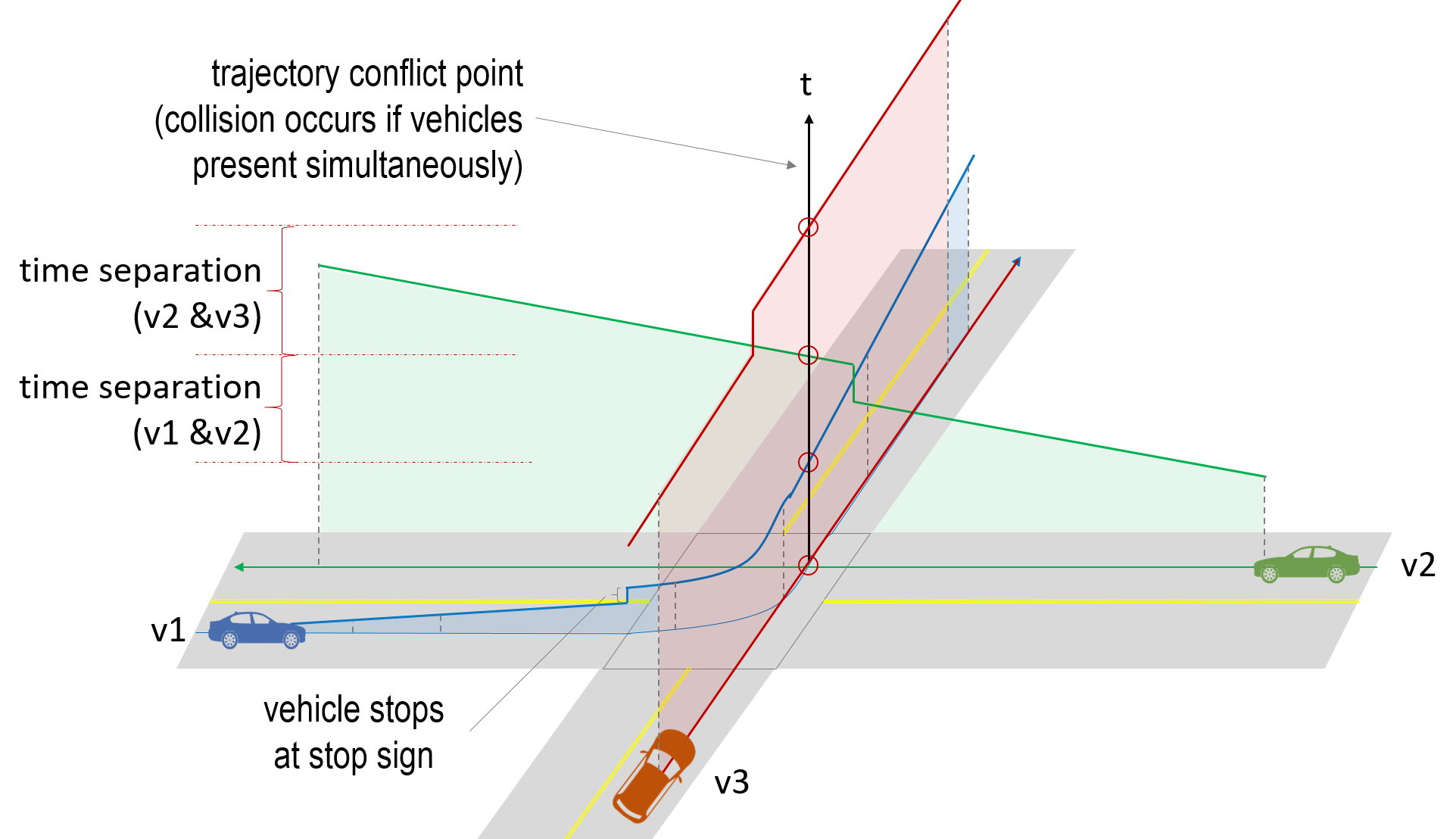}
	\caption{\textcolor{r1}{Vehicle-level conflict separation}} 
	\label{fig:ts_planning}
\end{figure}

The vehicle trajectory assignment is akin to aircraft separation planning in the time-space dimension. Fig. \ref{fig:ts_planning} visualizes the shared use of the finite intersection time-space resource of an all-way stop control (AWSC) intersection with a standard 4-leg layout. There are three vehicles crossing the intersection. Each line represents the trajectory of a individual vehicle. A crossing assignment is feasible as long as the trajectory lines (plus a safety buffer) do not intersect with each other. The separation in AIM is conducted at vehicle level with reservation-based systems, whereas the separation in SIM is ensured at the vehicle group level with traffic signals.

\begin{figure}[h]
\begin{minipage}[h]{.45\columnwidth}
\centering
\subfloat[intersection-based]{\label{main:a}\includegraphics[scale=0.13]{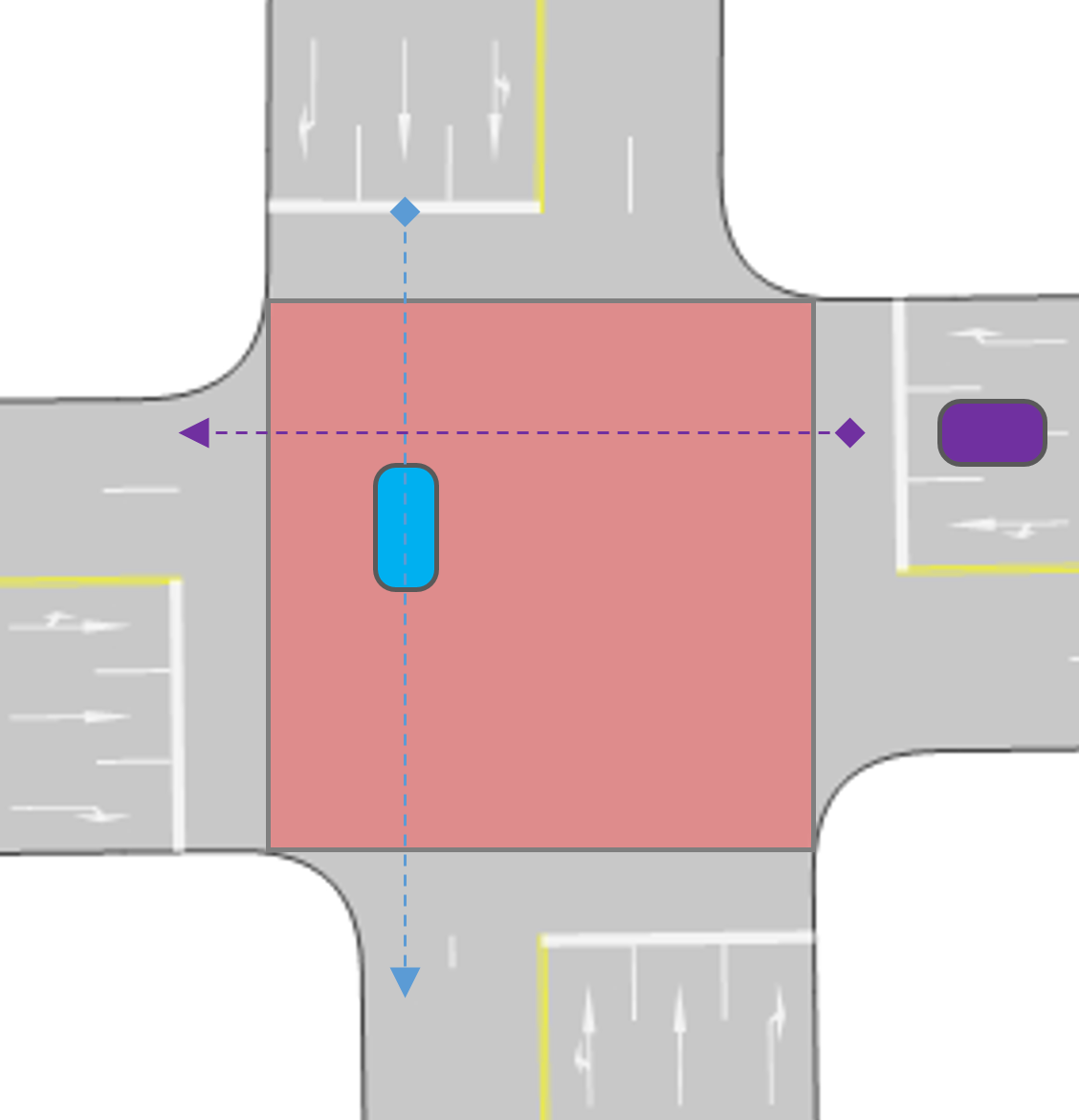}}
\end{minipage} 
\begin{minipage}[h]{.45\columnwidth}
\centering
\subfloat[tile-based]{\label{main:b}\includegraphics[scale=.13]{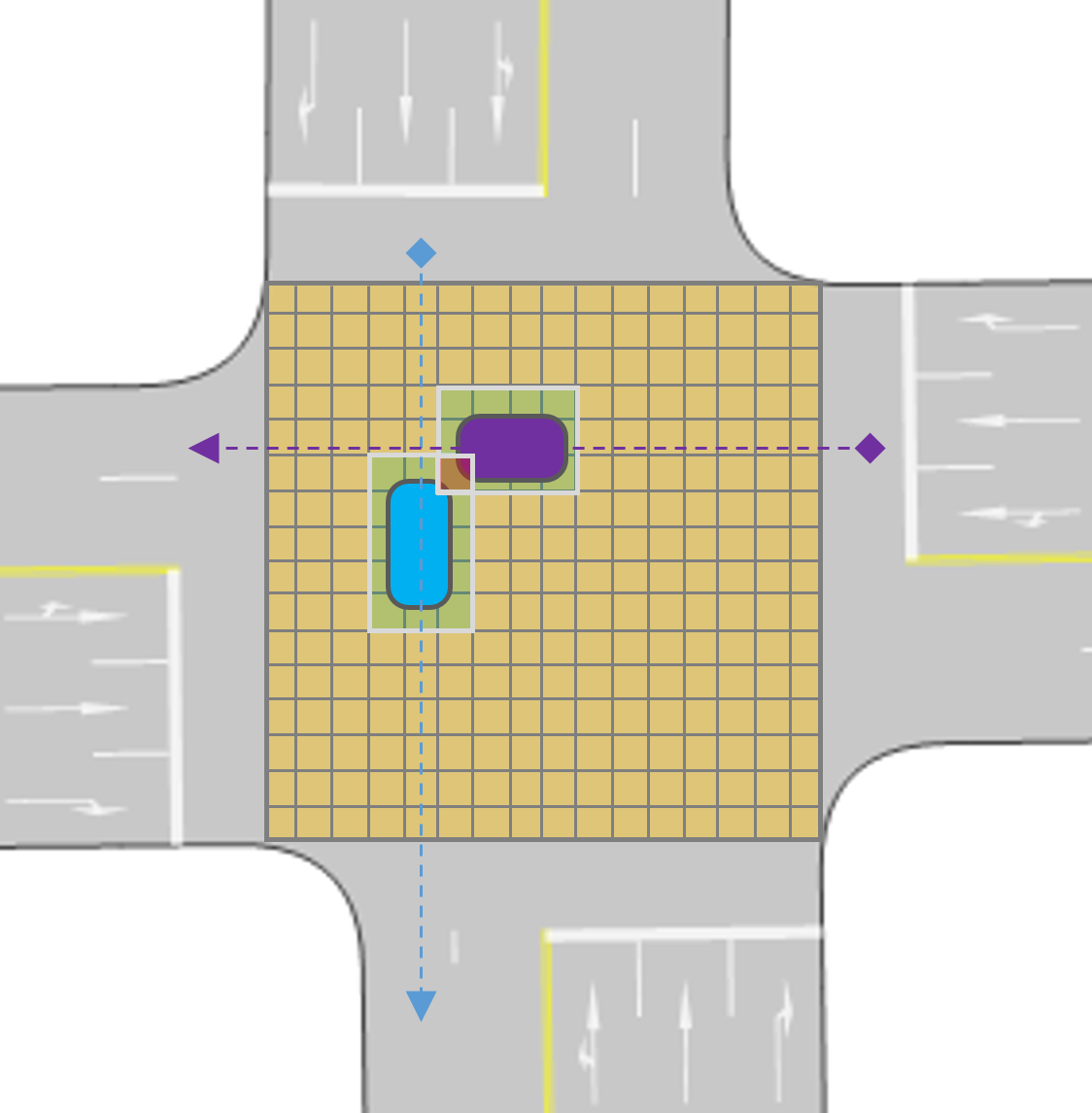}} 
\end{minipage}  \par
\begin{minipage}[h]{.45\columnwidth}
\centering
\subfloat[conflict point-based]{\label{main:c}\includegraphics[scale=.13]{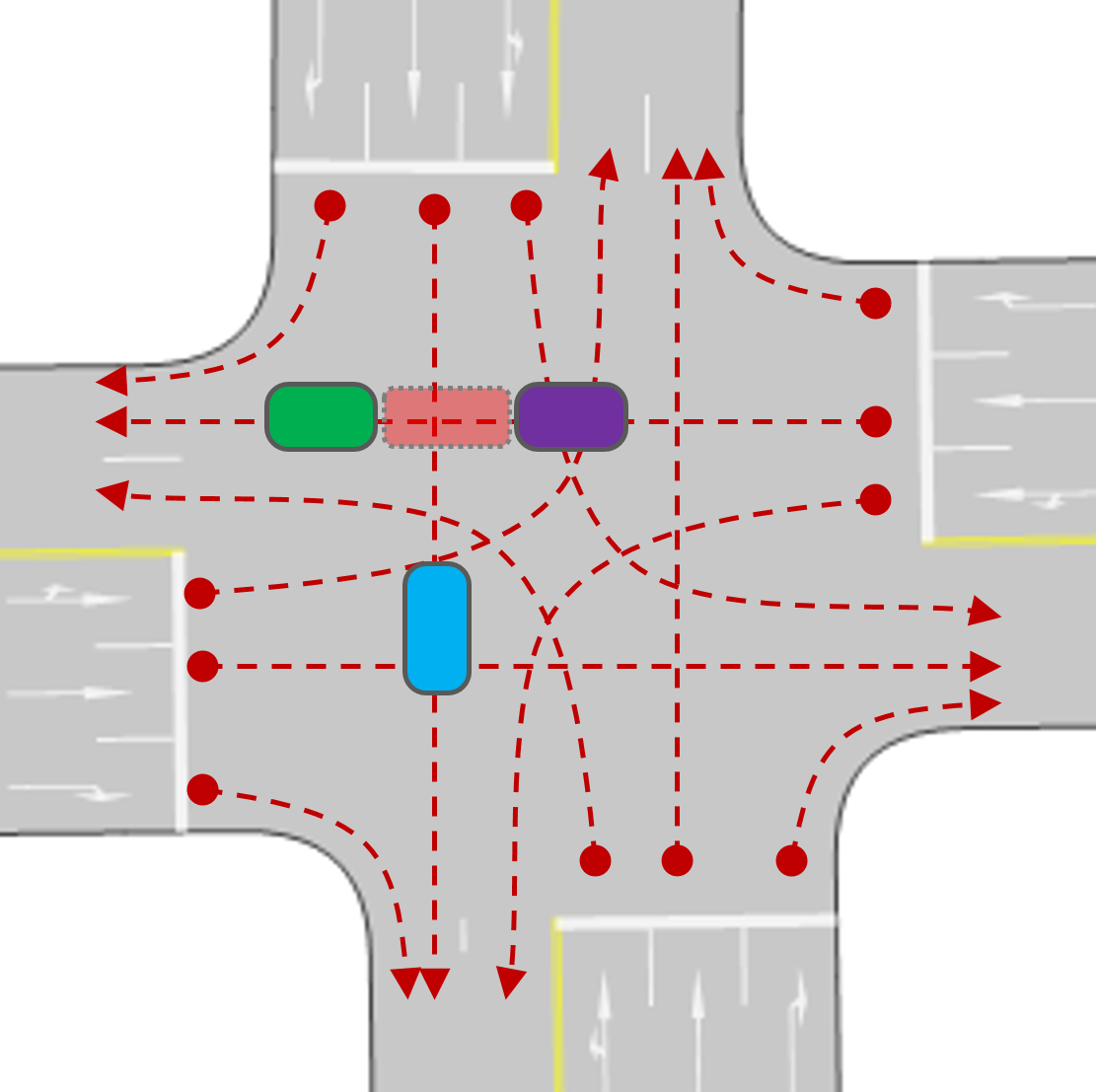}}
\end{minipage} 
\begin{minipage}[h]{.45\columnwidth}
\centering 
\subfloat[vehicle-based]{\label{main:d}\includegraphics[scale=.13]{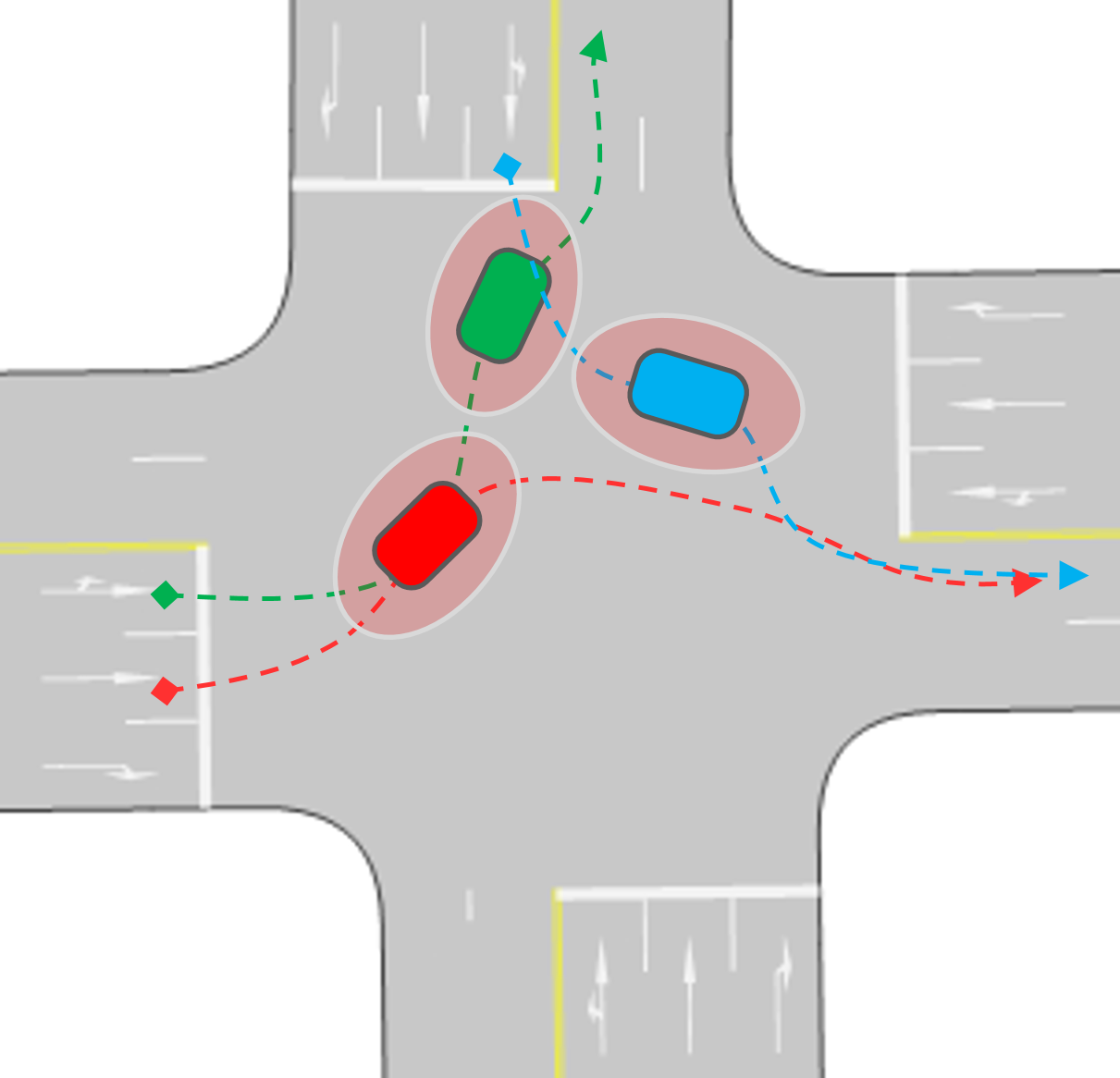}}
\end{minipage} 
\caption{Types of reservation for intersection management}
\label{fig:reserv_protocol}
\end{figure}

There are four reservation systems for separating conflicts in AIM: 
\begin{enumerate*}[label=\roman*)]
\item intersection-based reservation \cite{zhang2016optimal}, 
\item tile-based reservation \cite{dresner2008multiagent},
\item conflict point-based reservation \cite{kamal2015vehicle}, and
\item vehicle-based reservation \cite{Li2018near}
\end{enumerate*}. 
Intersection-based reservation allows one and only one vehicle within an intersection.
In tile-based reservation (shown in Fig. \ref{fig:reserv_protocol}(b)), space is discretized into a grid of tiles. A reservation is rejected if two vehicles occupy the same tile at the same time. Tiles could also be grouped into bigger regions to reduce the computation complexity for the reservation \cite{Bichiou2018b}. 
The conflict point-based reservation, shown Fig. \ref{fig:reserv_protocol}(c), is able to take full advantage of the intersection space \cite{kamal2015vehicle}. 
Li et al. \cite{Li2018near} proposed a radically different vehicle-based reservation system that is able to guide all the CAVs within a standard intersection (including the use of opposing travel lanes), provided collision is avoided. 
The vehicle-based reservation, as demonstrated in Fig. \ref{fig:reserv_protocol}(d), is the least restrictive reservation system; however, it imposes demanding computational expense to solve the nonlinear programming (NLP) problem due to the high-dimensional collision avoidance constraints. 

The trade-off between computational tractability and utilization of intersection space  was commonly tailored to the research need. Insofar, nearly all the studies for AIM dealt with standard intersections, in which there are three movements associated with each of the four approaches. Table \ref{table: reservation} shows the reservation systems in the previous studies. The majority of the AIMs included twelve (full) movements \cite{Ding2017a, muller2018optimal, liu2019trajectory, zohdy2016intersection}.  However, it is not uncommon to make simplifications by reducing the movements to only four \cite{Zohdy2012, medina2018cooperative, Bashiri2017, Fayazi2017, Lee2013}, or even two \cite{lam2013cooperative, muller2016intersection, jin2013platoon} during the concept development of an AIM.  To the best of our knowledge, no computational deadline was set for the trajectory planning among the reported studies, which means the AIMs were allowed to take as much time for solving the enter sequence. 

\begin{table*}[h]
\centering
\caption{AIM Reservation System}
\label{table: reservation}
\begin{tabular}{@{}lccccc@{}}
\toprule
\multirow{2}{*}{Literature} & \multicolumn{4}{c}{Reservation System} & \multirow{2}{*}{Movement} \\ \cmidrule(lr){2-5}
 & intersection-based & tile-base & conflict point-based & vehicle-based &  \\ \cmidrule(r){1-1} \cmidrule(l){1-6} 
 Fajardo et al. \cite{Fajardo2011} &  &  & \checkmark &   &  12\\
 Carlino et al. \cite{carlino2013auction} & \checkmark&  &  &  & 12  \\
 Bashiri and Fleming \cite{Bashiri2017} & \checkmark  &  &  &  &  n/a\\
 Bashiri et al. \cite{Bichiou2018} &  & \checkmark &  &  &  12\\
 Jiang et al. \cite{jiang2017distributed} & \checkmark &  &  &  & n/a\\
 Kamal et al. \cite{kamal2015vehicle} &  &  & \checkmark &  &  12\\ 
 M{\"u}ller et al. \cite{muller2016intersection} & \checkmark &  &  &  &  2\\
 Levin and Rey \cite{levin2017conflict} &  &  & \checkmark &  &  12\\
 Zohdy et al. \cite{Zohdy2012} &  & \checkmark &  &  &  4\\
 Du et al. \cite{du2018hierarchical} & \checkmark &  &  &  & 2 \\
 Ding et al. \cite{Ding2017a} &  &  & \checkmark &  &  12\\
 Bashiri et al. \cite{bashiri2018paim} &  \checkmark &  &  &  & 12 \\
 Li et al. \cite{Li2018near} &  &  &  & \checkmark &  unlimited\\
 Jin et al. \cite{jin2013platoon} &  &  & \checkmark &  & 2  \\
 Mirheli et al. \cite{mirheli2018development} &  &  & \checkmark &  &  8 \\
 Li and Zhang\cite{Li2018a} &  &  &  & \checkmark   &  unlimited\\
 Wuthishuwong et al.\cite{wuthishuwong2015safe} &  & \checkmark &  &  &  12\\
 Liu et al. \cite{liu2019trajectory} &  & \checkmark &  &  &  12\\
 Zohdy and Hakha \cite{zohdy2016intersection} &  &  & \checkmark &  & 12  \\
 Zhao et al. \cite{Zhao2018multi} &  &  & \checkmark   &  &  12\\
 Hassan and Hakha \cite{hassan2014fully} & \checkmark  &  &  &  &  4\\
 Stone et al. \cite{stone2015autonomous} &  &  &  &  &  12\\
 Fayazi et al.\cite{Fayazi2017} && & \checkmark & &4\\
 Lam and Katupitiya \cite{lam2013cooperative} & \checkmark & & & & 2\\
 Creemers et al. \cite{creemers2018design} &  & &\checkmark  & &  5\\
 \bottomrule
\end{tabular}
\end{table*}

\subsection{Priority Policy}
The priority policy dictates the allocation of the intersection resources. It is independent of the reservation framework, whose main objective is to separate conflict movements. In SIM, the priority among vehicle groups is determined by operational needs (e.g., queue length or delay) and implemented via the SPaT plan. Priority can be increased by extending the green signal timing for movement groups of interest.
 
Different from SIM, the priority assignment of AIM is on an individual vehicle level. The only vehicle-level priority assignment on SIM is transit- or emergency- vehicle signal preemption. However, it only makes exceptions for a few special vehicles. When it comes to AIM, the first-come-first-serve (FCFS) policy, which is based on fairness, has been adopted in the majority of the AIM research. The system-optimal policy is the second most-used priority policy where the crossing sequence is determined based on system-level performance measures, such as overall delay, throughput, travel time, etc. Other priority policies have been reported, such as longest-queue-first policy\cite{wu2019dcl}, vehicle type-based policy \cite{dresner2006human}, custom priority score-based policy \cite{liu2019trajectory}, and auction-based policy \cite{carlino2013auction}.

Table \ref{table:aimLayerCoverage} lists the major AIM study with regard to the coverage of the three-layer intersection management structure. Few studies of AIM investigated the corridor-level coordination of AIMs. An increasing amount of AIM studies that coupled an explicit vehicle dynamics model was observed. Game-theoretic priority policy \cite{Zohdy2012, Elhenawy2015} is a popular choice among the heuristic methods.  Platoon-based performance metrics \cite{Bashiri2017,bashiri2018paim} is another type of popular heuristic methods.

\begin{table*}[]
\centering
\caption{Layer Coverage of AIM Studies} 
\begin{tabular}{lccccccc}
\hline
\multirow{2}{*}{Study} & \multirow{2}{*}{Reservation} & \multicolumn{3}{c}{Layer} & \multicolumn{3}{c}{Priority} \\ \cline{3-8} 
 &  & corridor & intersection & vehicle & FCFS & system optimal & heuristic \\
\hline
Jiang et al. \cite{jiang2017distributed} & IB &  &  & \checkmark & \checkmark &  &  \\
Bichiou and Rakha\cite{Bichiou2018b} & TB  &  & \checkmark & \checkmark &  & \checkmark \\
Liu et al.\cite{liu2019trajectory} & TB &  & \checkmark &  &  &  & \checkmark   \\
Ding et al.\cite{Ding2017a} & CP &  & \checkmark &  &  & \checkmark &  \\
Fajardo et al.\cite{Fajardo2011} & CP &  & \checkmark &  & \checkmark   \\
Hassan and Hakha\cite{hassan2014fully} & IB &  &  &  &  &  & \checkmark  \\
Kamal et al.\cite{kamal2015vehicle} (VICS) & CP &  & \checkmark & \checkmark &  & \checkmark & \\
Stone et al.\cite{stone2015autonomous} & TB+SIM  &  &  &  & \checkmark &  &  \\
Levin and Rey \cite{levin2017conflict} (CPIC-AIM) & CP &  & \checkmark & \checkmark &  & \checkmark & \\
Lam and Katupitiya\cite{lam2013cooperative} & IB &  &  & \checkmark &  &  & \checkmark  \\
Li et al.\cite{Li2018near} & VB &  & \checkmark & \checkmark &  &  & \checkmark \\
Du et al.\cite{du2018hierarchical} & IB & \checkmark & \checkmark & \checkmark &  & \checkmark \\
Jin et al.\cite{jin2013platoon} & CP &  & \checkmark &  &  &  & \checkmark \\
Zhao et al.\cite{Zhao2018multi} & CP &  &  & \checkmark & \checkmark &  &  \\
Bashiri and Fleming\cite{Bashiri2017} & IB &  & \checkmark &  &  &  & \checkmark  \\
Bashiri et al.\cite{bashiri2018paim} & IB &  & \checkmark    &  &  &  & \checkmark \\
M{\"u}ller et al.\cite{muller2016intersection} & IB &  & \checkmark &  &  & \checkmark &   \\
Wuthishuwong et al.\cite{wuthishuwong2015safe} & CP &  &  \checkmark &  &  &  &\checkmark    \\
Mirheli et al.\cite{mirheli2018development} & CP &  & \checkmark &  &  & \checkmark &  \\ 
Carlino et al.\cite{carlino2013auction} & IB & \checkmark & \checkmark &  &  &  & \checkmark \\
Creemers et al.\cite{creemers2018design} & CP &   & \checkmark  & \checkmark   &   & \checkmark  \\
\hline
\end{tabular}
\label{table:aimLayerCoverage}
\end{table*}

\subsubsection{Fairness-based Priority Policy}
The FCFS policy has been widely adopted in existing studies. First-in-first-out (FIFO) is an alternative term that has been used. The FCFS sequence is determined by two criteria:
\begin{enumerate*}[label=\roman*)]
\item the estimated arrival time to the infrastructure-to-vehicle (V2I) communication boundary of the intersection (i.e., stop line), 
\item the arrival of the communication boundary.
\end{enumerate*}
\textcolor{r1}{If the identical arrival times of multiple vehicles are identified, the right-of-way could be assigned to the vehicle on this right (in right-hand driving countries). This rule has been in practices to resolve the conflict arisen from AWSC intersection when two vehicles arrive at the stop sign at the same time, albeit more sophisticated rule is certainly possible for AIM.}

Algorithm \ref{alg:MZScheduler} generalizes the conflict point-based FCFS reservation system. The algorithm is also applicable to non-standard intersection layouts, for instance the diverging diamond interchange (DDI), as long as the conflict points are identified. $N$ is the total number of vehicles that need to cross an intersection; $t_{k}^{c2}$ is the time at conflict point $c2$ for vehicle $k$; $\delta$ is minimum distance; $S$ is the distance to be traversed withing the intersection; $t_{j}^{in} $ is the time vehicle $j$ enters the intersection; $t_{j}^{out} $ is the time vehicle $j$ exits the intersection; $t_{j,k}^{in}$ is the enter time for vehicle $j$ considering the conflict with vehicle $k$.
\begin{algorithm}
\caption{Conflict point-based FCFS-AIM scheduler}
\label{alg:MZScheduler}
\begin{algorithmic}[t]
\State \textbf{initialization}  \Comment{get intersection info. (e.g., conflict points)}
\State get the intersection entry sequence 
\For{\texttt{$1<j<N$}}
    \For{\texttt{$k<j$}}
        \State identify conflict point $c_{j,k}$ between vehicle $j$ and $k$, $c_{j,k} \in \left \{1, 2, \cdots , Q_{j}\right \}$
    \State obtain the entry time $t_{k}^{in}$ of vehicle $k$  
    \State $t_{j,k}^{in} = t_{k}^{Q} + \frac{S}{v_{intx}} + \delta$    \Comment{assuming $Q$ is the conflict point}
    \EndFor
    \State $t_{j}^{in} = max \left \{ t_{j,1}^{in}, t_{j,2}^{in},\cdots, t_{j,k}^{in}\right \}$  \Comment{get the most conservative entry time}    
    \State subsequently update and store $t_j^{in},t_{j}^{1},t_{j}^{2}, \cdots t_j^{Q_j}, t_j^{out} $
\EndFor
 \label{Algorithem}
\end{algorithmic}
\end{algorithm}

FCFS is able to achieve good performance under certain circumstances. Fajardo et al. \cite{Fajardo2011} compared the FCFS-based AIM with SIM and found that the FCFS protocol significantly outperformed SIM in various testing scenarios. The comparison was among three FCFS-AIMs (with combinations of a static buffer, internal time buffer, and edge time buffer, respectively) and SIM (with single protected left-turn phase) under low, medium, and high volume scenarios. 

There are several issues with the FCFS policy. First, the FCFS may impose an external cost for other vehicles with higher priority due to its priority-agnostic nature. Imagine an extreme case where an emergency vehicle is at the back of the queue, waiting to clear the intersection, such vehicle is only granted permission to enter the intersection after all the preceding vehicles in the FCFS queue. 
\textcolor{r1}{The handling of emergency vehicle was reported in a handful of studies. Dreser and Stone \cite{dresner2006human} proposed the FIFS-EMERG AIM, an augmentation of existing FIFS-AIM that grants priority to the lane of the incoming emergency vehicle. The lane-level priority is implemented to ensure that the non-emergency vehicles do not stop on the travel lane, which could potentially block the emergency vehicle. It was found that the FIFS-EMERG yielded lower average delay for emergency vehicles. The handling of emergency vehicle is relatively straight forward for non-FIFS-AIM, such as by increasing emergency vehicle's priority score \cite{He2012} or giving it a virtually infinite budget in the auction-based AIM \cite{carlino2013auction}
}

Second, a reservation is meaningful only if the requesting vehicle is able to execute it. In another words, a vehicle in a queue may not request a reservation until it is able to enter an intersection. This suggests that an intersection approach with more lanes is likely to obtain a greater share of the intersection capacity as more vehicles in the front of the lanes can request reservations at the same time. Furthermore, the FCFS policy assigns equal weight to all approaches, which means a vehicle on a minor approach can break the progression of vehicle platoons on the major approach. Third, FCFS does not strictly maintain the order of entry and the order of reservation requests. For instance, let $i, j, k$ be the indexed vehicles in the FCFS queue in ascending order. Vehicle $i$ obtains a reservation, whereas vehicle $j$ got rejected due to conflict with vehicle $i$, but vehicle $k$ was accepted in the absence of conflicts with vehicle $i$ and $j$. In this case, we have an entry sequence of $[i, k, j]$, different from the FCFS order $[i, j, k]$. 

Levin et al. \cite{levin2016paradoxes} presented a theoretical example of the exploitation of the FCFS policy and demonstrated the superiority of SIM to FCFS-AIM. The simulation of an arterial network (with 5 signalized intersections and 21 links) revealed that the AIM was outperformed by a traffic signal in all demand levels with the exception of under low demand scenarios. The FCFS-AIM was subsequently evaluated on a large-scale urban network (i.e., downtown Austin, TX), where all SIMs were replaced with FCFS-based AIM.  With the additional assumption of user equilibrium route choice, interestingly, the superior performance of FCFS-based AIM was observed in the urban grid network: the overall travel time decrease by over 50\% in all scenarios.  Levin et al. concluded that it was the availability of parallel links in an urban network, combined with the user equilibrium route choice, that evenly distributed traffic to avoid high delay intersections, despite the theoretical disadvantages of the FCFS policy.  

A probabilistic model based on the turning ratio of a standard intersection was put forward to theoretically compute the saturation flow rate of a conflict point-based AIM. The capacity was found to be 1,667 vehicle per hour (vph) assuming no turning movements, compared to the capacity range of 1700-1900 vph for SIM \cite{manual2016highway}. 
Zhang and Cassandras \cite{zhang2018penetration} compared the FCFS-AIM with decentralized optimal control at vehicle level with SIM. They concluded that as traffic grows, a higher CAV penetration is necessary to match the performance of SIMs.  $750$ vph per lane was considered as the critical flow rate. When the traffic demand is above 750 vph per lane, even assuming 100\% CAV penetration, FCFS-AIM still cannot outperform SIM in terms of energy saving. Under saturated condition, nearly all vehicles have to slow down or even stop to create the necessary separation for entering the intersection. However, non-signalized coordination is more effective in reducing travel delays than signalization.

\subsubsection{System-optimal Priority Policy}
FCFS is not likely to produce the system optimal solution, as it does not explicitly optimize the global intersection performance  \cite{zohdy2016intersection}. A trajectory management layer for the entire intersection can be added to relax the FCFS policy.  Lee and Park \cite{Lee2013} put forward a cooperative vehicle intersection control (CVIC) system which did not require any signalization even under moderate intersection demand (1,900 vph). In CVIC, each passing vehicle with potential conflict with another vehicle was assigned with an individual trajectory to minimize the trajectory overlapping within an intersection. A general trajectory planning for minimizing the overlapping of vehicle trajectories is shown in (\ref{eq:trajPlanningFramework}). 
\begin{subequations}
\label{eq:trajPlanningFramework}
\begin{gather} 
\min_{v(t)}~ J=\sum_{\phi=1}^{P}\sum_{l=1}^{L} \sum_{n=1}^{N}\int_{q}^{p}\sqrt{(1+ v_n(t))^2)dt}\\
\text{subject to: }\\  
a_{n,min}  \leq a_{n}(t)\leq a_{n,max},\quad\\
0 \leq v_{n, min}\leq v_{n}(t)\leq v_{n, max},\quad\\    
\tau < x_{n}(t) - x_{n+1}(t)
\end{gather}
\end{subequations}
where, $P$ is the total number of the movement types; $\phi$ is the movement type index; $L$ is the total number of lanes of movement $\phi$ with lane index $l$;  $N$ is the total number of vehicles; $n$ is the vehicle index; $v_n(t)$ is the time-dependent speed of vehicle $n$; $p$ is the arrival time at the beginning of intersection; $q$ is the exiting time at the end of the intersection; $v_{n, max}$ is the maximum speed; $v_{n, min}$ is the minimum speed;  $a_{n, max}$ is the maximum acceleration; $a_{n, min}$ is the minimum acceleration; $x_{n}(t)$ is the position of vehicle $n$; $\tau$ is the minimum headway on consecutive vehicle on the same lane.

A centralized cooperative intersection control was proposed by Ding et al. \cite{Ding2017a}. The control strategy was designed to minimize intersection delay, fuel consumption as well as emission, while avoiding a collision by separating the arrival time for conflict vehicle at each conflict point. Fayazi et al. \cite{Fayazi2017} formulated the intersection management as a mixed-integer linear program (MILP) to maximize the number of vehicles that clear the intersection within a given interval and, at the same time, minimize the difference between desired arrival time and assigned time. A signal-free intersection control logic (SICL) was proposed in \cite{mirheli2018development} to maximize the intersection throughput. Dynamic programming was employed to find the near-optimal trajectory under safety constraints. Later the framework was improved with a cooperative vehicle-level structure to account for CAV preferences and scalability \cite{mirheli2019consensus}.
Ding et al. \cite{Ding2017a} proposed a multi-objective optimization model for minimizing delay, emission, and discomfort level. Kamal et al. \cite{kamal2015vehicle} proposed a vehicle-intersection coordination scheme (VICS), which used a risk score as the objective. The risk score quantitatively indicated the potential risk of collision at a time step for a vehicle pair based on the overlapping area of the two-dimensional Gaussian functions.

\subsubsection{Heuristic Priority Policy}
\label{subSecPriorityPolicy}
\begin{figure} [h]     
	\centering
	\includegraphics[width=\columnwidth]{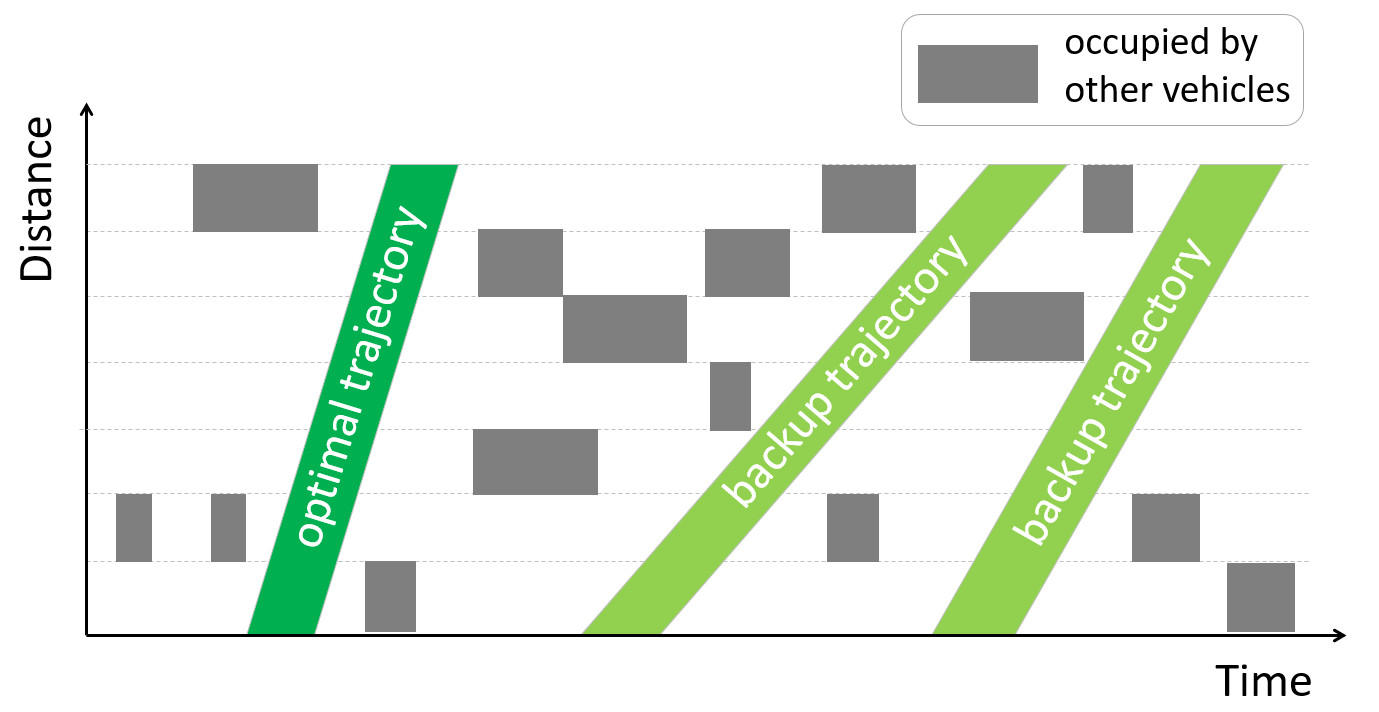}   
	\caption{Window searching for a vehicle going straight}
	\label{fig:winSearch}
\end{figure}

To account for service priorities, Liu et al. \cite{liu2019trajectory} proposed an intersection management framework called TP-AIM. Here, a window searching algorithm (illustrated in Fig.  \ref{fig:winSearch}) was proposed to find the entry window that yields a collision-free trajectory, while factoring in the service priority (e.g., emergency vehicles, heavy-duty trucks, school buses) as well as the vehicle-based score. The vehicle-based score was determined by the distance to the intersection, headings, etc. The assignment for vehicles with lower service priority had to first take into account the vehicles with higher service priority.

Game-theoretic approaches have been incorporated into AIM. To put into a transportation perspective, all CAVs could potentially form a cooperative game along with the intersection controller via V2X communication. Elhenawy et al. \cite{Elhenawy2015} proposed a game-theory-based algorithm, where CAVs communicate vehicle status (i.e., speed and location) to a centralized intersection manager. In the proof-of-concept simulation, two sets of vehicles (north-sound and east-west) were classified as \textcolor{r1}{two} players, each of whom tried to minimize their delay at the intersection. Each player had three options: to accelerate, to decelerate, and to maintain current speed. Upon obtaining vehicle information, the intersection manager solves the game matrix and obtains the Nash equilibrium. Then, the optimal actions were distributed to each vehicle. Compared to an AWSC intersection, the proposed scheme achieved 49\% and 89\% reduction in vehicle travel time and delay, respectively. A CACC-CG (Cooperative Adaptive Cruise Control - Cooperative Game) was proposed in \cite{Zohdy2012}. The CACC-CG is comprised of a manager agent and reactive agents at each time step. The manager agent selected one reactive agent for movement optimization. The reactive agents, using symmetric information that is shared among players, choose among acceleration, deceleration, or maintaining current speed.  All the players choose the minimum utility value from the payoff table at each time step to achieve an equilibrium state.  

The platoon-based reservation has steadily gained recognition. In \cite{Bashiri2017}, the benefits of forming platoon among crossing vehicles were studied. The study proposed using platoon leaders to communicate on-behalf of followers to decrease communication complexity. The Platoon-based Delay Minimization cost function and the Platoon-based Variance Minimization cost function were formulated for scheduling the crossing sequence. In a subsequent study, Bashiri and Fleming \cite{Bashiri2017} introduced a reservation policy that minimized delay or optimized schedules. To tackle the exponential nature of the permutation on the crossing sequence (i.e., ($O(N^{N}$)), a heuristic method that ignores the non-conflicting trajectories, was proposed to reduce the computation complexity to $O(N!)$.

\subsection{Centralization}
\begin{figure*}[h]
\begin{minipage}[h]{.65\columnwidth}
\centering
\subfloat[centralized AIM]{\includegraphics[scale=0.17]{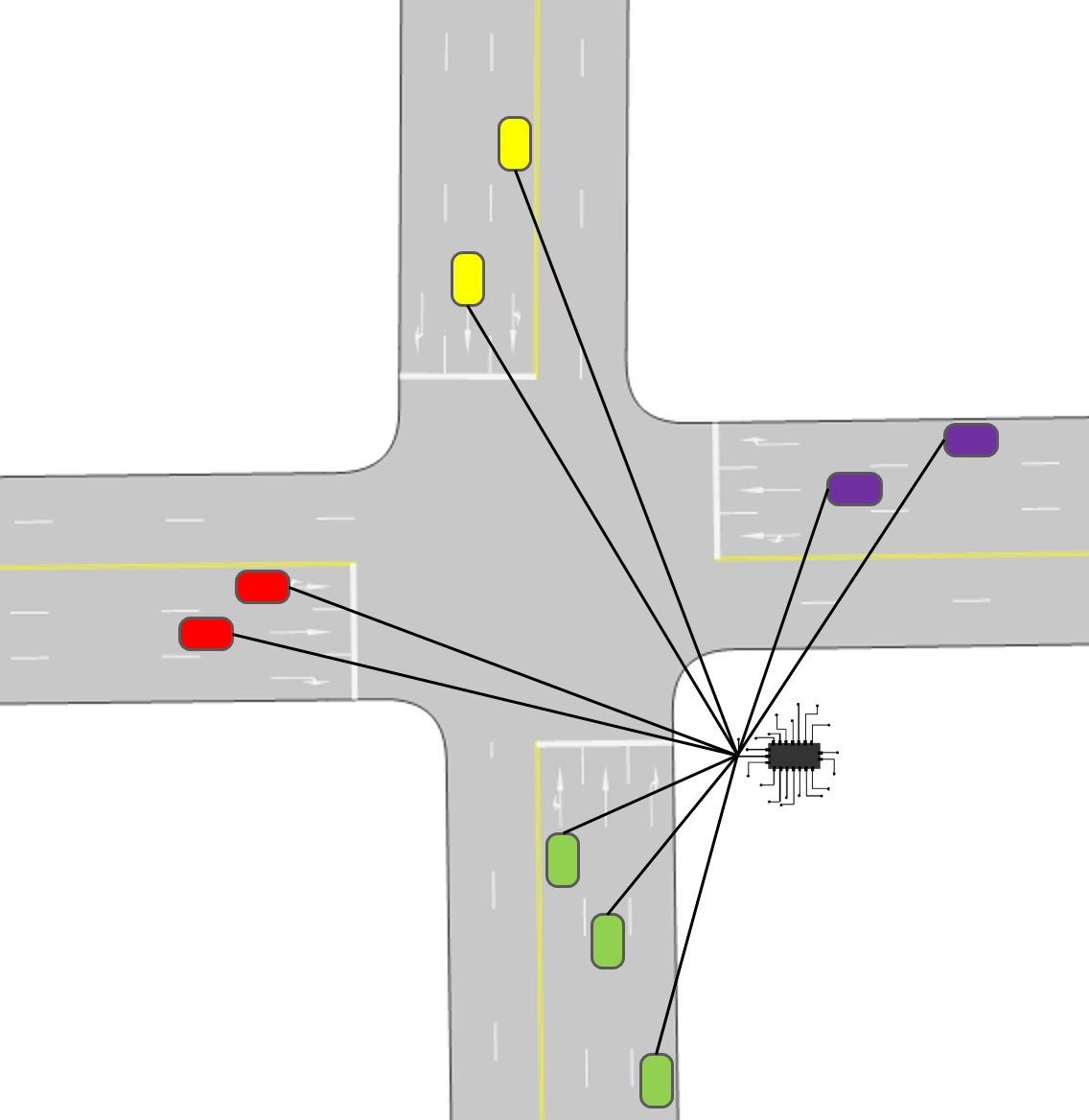}}
\end{minipage} 
\begin{minipage}[h]{.65\columnwidth}
\centering
\subfloat[decentralized AIM]{\includegraphics[scale=.17]{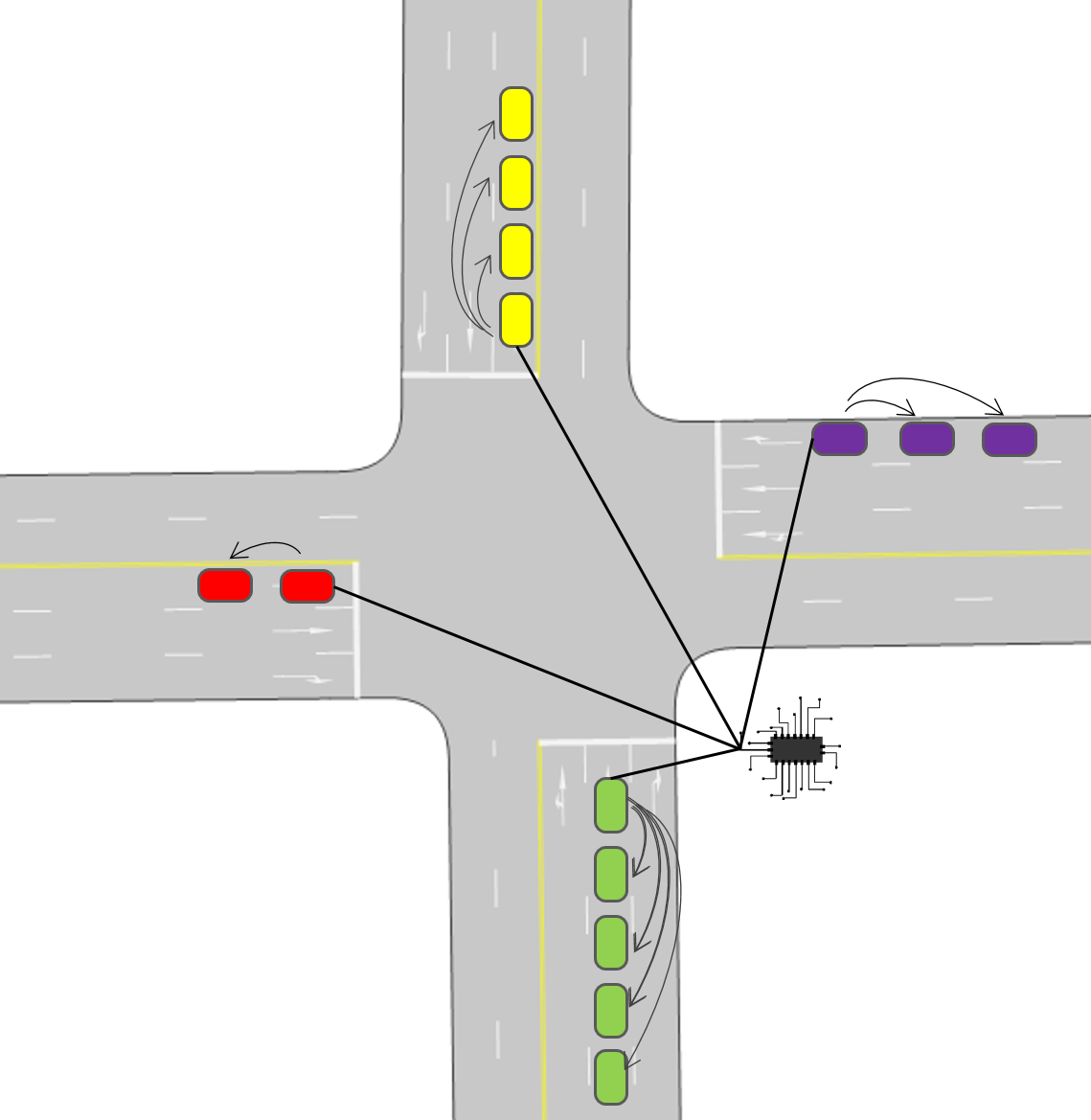}}
\end{minipage} 
\begin{minipage}[h]{.65\columnwidth}
\centering
\subfloat[distributed AIM]{\includegraphics[scale=.17]{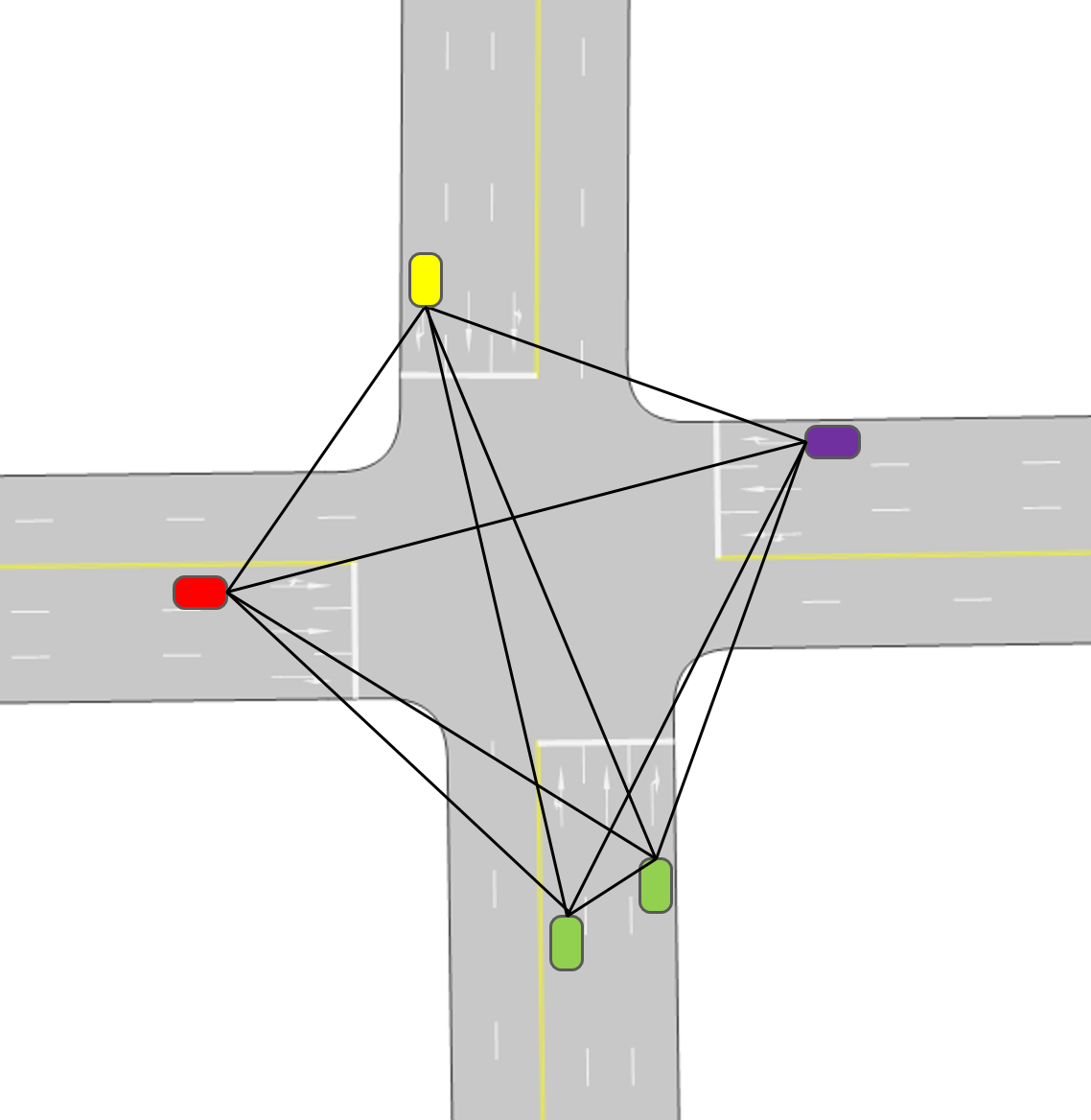}}
\end{minipage}
\caption{Communication structure of of AIM}
\label{fig:centralization}
\end{figure*}

``Centralized", ``decentralized", and ``distributed" are the terms used for describing the organization of AIM systems.  The information flow and the corresponding organization of the three types of planning is illustrated in Fig. \ref{fig:centralization}. 


\textbf{Centralized AIM} has a single-point contact among nodes (vehicles) for information sharing and decision making, as shown in Fig. \ref{fig:centralization}(a). As such, single-point failure is the primary concern for centralized AIM. Early AIMs relied on centralized control strategies where the intersection manager guided the CAVs to safely traverse the intersection. Centralized intersection management strategies are costly to implement, and their scalability is open to question \cite{hassan2014fully}. The current state of V2X wireless communication may not technologically guarantee such performance with thousands of vehicles in the vicinity of an intersection. 

\textbf{Decentralized AIM} contains several central hubs within the systems. Note that the processing in the distributed system is shared across multiple nodes. For instance, for the platoon-based AIM \cite{Bashiri2017,bashiri2018paim}, the platoon leader acts as the decentralized hub to communicate with the intersection manager to obtain permission to enter the intersection. The intra-platoon communication for platoon following is assumed to conduct locally among the platoon members as demonstrated in Fig. \ref{fig:centralization}(b). 

\textbf{Distributed AIM} is an extreme case for decentralization. In a distributed system, as exhibited in  Fig. \ref{fig:centralization}(c), there is not a single point where the decision is made, and each node makes a decision for its own behavior. The system behavior is the result of the aggregated response for each node within the system. 
The distribution of the scheduling among vehicles has the potential of becoming truly fault-tolerant. Hassan and Rakha \cite{hassan2014fully} proposed a fully-distributed heuristic intersection control strategy that aims to minimize the communication (information exchanges) in each time step.  The vehicles approaching the intersection are categorized into four groups (``Out,'', ``Last'', ``Mid'', and ``Head''), the group closest to the intersection (the ``Head'' group) assumes the role of the schedulers, which are responsible for the passages of the intersection of all vehicles at different non-conflicting times.

\subsection{Summary}
The trajectory planning for ensuring safety is the core function of AIM. The vehicle-level crossing assignment allows a great deal of flexibility, but at the same time increases the search space of entry sequence drastically. Trajectory planning is usually formulated as a nonlinear, non-convex problem \cite{zhu2015linear} in order to fully satisfy the collision avoidance requirement. Mixed-integer programming (MILP) has been seen in many formulations of AIM \cite{Fayazi2017, levin2017conflict, He2012, mirheli2019consensus}.  The FCFS trajectory planning provides a simple and fair way to gain intersection access. However, its efficiency is open to debate. Efforts have been made to factor in the priority of different vehicles.

\section{Vehicle Control for AIM}
\label{sec:VehControl}
AIM research from the traffic engineering perspective typically assumes the availability of the vehicle control and emphasizes on coordinating conflicting crossing movements. The roadside unit (RSU) take over the control of the vehicle and guide it to safely cross the intersection. Intersection control, coupled with vehicle control from control engineering, is a promising and necessary direction in AIM research. 
With vehicle automation, the driving function is expected to be replaced by the vehicle controller which have been actively studied from control engineering perspective. As shown in Fig. \ref{fig: controlLayer}, the control of individual vehicle is classified as the lowest layer in the hierarchical framework for AIM.  A second-order dynamics model is typically used for vehicle control in relevant AIM studies, as expressed in (\ref{eq:state}), \textcolor{r1}{where $p_i(t)$ is the position of vehicle $i$ at time $t$ and $v_i(t)$ is the speed of vehicle $i$ at time $t$  }
\begin{equation} \label{eq:state}
\dot{p_i}(t) = v_i(t), \dot{v_i}(t)=a_i(t) 
\end{equation} 

\begin{figure} [H]
	\centering
	\includegraphics[width=0.8\columnwidth]{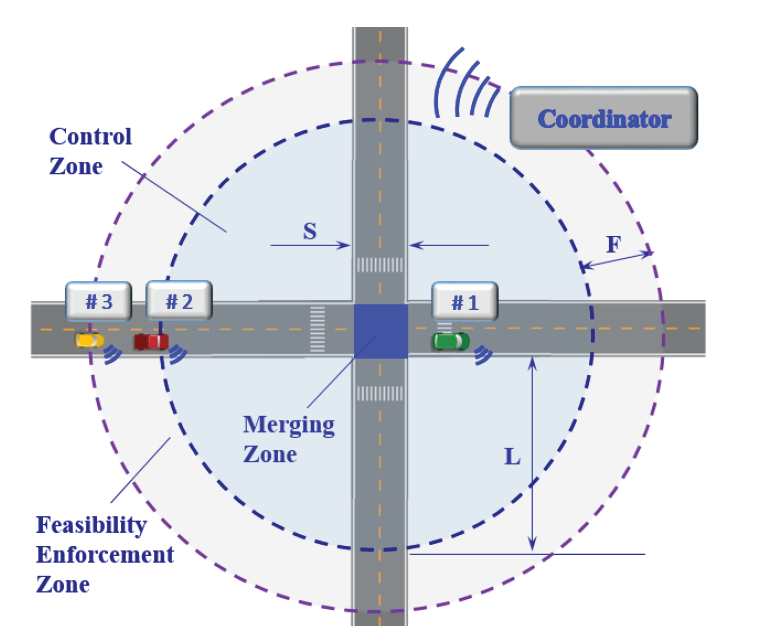}   
	\caption{Intersection zones (source: \cite{zhang2017optimal})} 
	\label{fig:inxtZone}
\end{figure}

\subsection{Optimal Control}
The underlying concept of optimal control is to find a control strategy that yields minimum cost for the associated process while satisfying the applicable control and state constraints \cite{tonon2017optimal}. A basic optimal control framework for a CAV is expressed in \ref{eq:controlFramework}, which aims at minimizing the performance index $J$ \textcolor{r1}{(a.k.a., cost function)}. The configuration of the performance index $J$ influences the optimal control of a vehicle. The common performance indexes include control input $u_i^2(t)$, acceleration $a_i^2(t)$, jerk $\dot{a}_i^2(t)$, derivative of jerk $\ddot{a}_i^2(t)$, and evacuation time (expressed as $\int_{t_i^0}^{t_i^f} dt$).  
\begin{subequations}
\begin{gather} 
\min_{u_i(t)}~ J=\int_{t_i^0}^{t_i^f} U(x, t) dt\\
\text{Subject to: }(\ref{eq:state}),\\  
u_{i,min}  \leq u_{i}(t)\leq u_{i,max},\quad \\
a_{i,min}  \leq a_{i}(t)\leq a_{i,max},\quad \\
0 \leq v_{min}\leq v_{i}(t)\leq v_{max},\quad \forall t\in\lbrack t_{i}%
^{0},t_{i}^{f}],\\
\text{and given}~t_i^0, t_i^m,t_i^f, p_i(t_i^0)=0, p_i(t_i^f)=L, v_i(t_i^0), u_i(t_i^0), \nonumber
\end{gather}
\label{eq:controlFramework}
\end{subequations}
where $t_i^0$ is the entry time for vehicle $i$ to the control zone (Fig. \ref{fig:inxtZone}); $t_i^m$ is the entry time to the intersection; $t_i^f$ is the time for clearing the intersection; $u_{i}(t)$ is the control input for vehicle $i$ at time $t$; $p_i$ is the location in the control zone with length $L$; all other variables are as previously defined.

Decentralized optimal vehicle coordination was proposed for intersection management by Jiang et al. \cite{jiang2017distributed}. Under this framework, each vehicle solves its own optimal control problem and then exchanges arrival and exiting time with neighboring vehicles. The entry sequence of the vehicles was assumed to be available. With ten vehicles, the non-convex problem can be reliably solved. However, the algorithm may not scale well in high volume scenarios as more collision avoidance constraints may become active, making the problem computationally intractable.

The co-design of optimal vehicle controls and crossing scheduling for intersection is complex with little available methods \cite{Yao2018}.  The majority of the studies used an upper intersection management layer to assigned a collision-free entry sequence to the intersection. Zhang et al. \cite{zhang2016optimal} proposed an optimal intersection control framework with intersection-based FCFS policy. An intersection manager scheduled the entry time of each CAV, which separated the conflicts among different movements within the intersection. With the assigned entry time, a vehicle then executes the optimal control strategy, which is derived by Hamiltonian analysis \cite{naidu2002optimal} with the assumption that none of the constraints was active within $[t_i^0, t_i^m]$. The consideration of left and right turning movements was introduced in \cite{zhang2017decentralized}.  The AIM framework considering turning movements with optimal control was presented in Algorithm \ref{alg:optmal_control_framework}. Their system solves the control strategy for an individual vehicle, but it did not implicitly coordinate the scheduling among different conflicting movements. The FCFS sequence may need to update depending on the movements of consecutive vehicles set in the FCFS queue.

Zhang et al. \cite{zhang2017optimal} later proved the existence of a non-empty set of initial conditions that keep the collision avoidance constraint inactive over the entire control zone. Hence the optimal control strategy is theoretically attainable.  To ensure the feasible initial condition, they proposed a feasibility enforcement zone located upstream of the control zone in Fig.\ref{fig:inxtZone}. The constrained optimal control was addressed by Malikopoulos et al. \cite{malikopoulos2018decentralized}, where the constrained and unconstrained arcs were pieced together according to the activation of one or multiple constraints. First, the time when the control (or state constraint, or both) become active was determined. Then, the constraint and non-constraint arcs needed to be pieced together. For two state constraints and two control constraints, there are six cases for constructing the constrained arcs.

\begin{algorithm}
\caption{Decentralized Optimal Control Framework for AIM  (source: \cite{zhang2017decentralized})}
\label{alg:optmal_control_framework}
\begin{algorithmic}[h]
\State initialize intersection info 
\For{\texttt{vehicle $j<N$}}  \Comment{loop through FCFS queue}
    \For{\texttt{$k<j$}}
        \State assign vehicle $k$ into one of the four pre-determined conflict sets (i.e., $\mathcal{E}_i,S_i,L_i, O_i $ ) based on intersection movements
		\begin{itemize}    
	\item $\mathcal{E}_i$ with all the vehicles that can cause rear-end collision at the end of the intersection
    \item $S_i$ with vehicles on the same lane that can cause rear-end collision at the beginning of the intersection
    \item $L_i$ with vehicles  with different origin-destination that can cause collision within the intersection
    \item $O_i$ with vehicles with different origin-destination that cannot cause collision within the intersection 
		\end{itemize}
        \State retrieve the terminal time $t_k^f$ and entry time $t_k^m$ for the closest preceding vehicle within each set
    \EndFor
    \State determine the terminal time $t_j^f$ from each of the conflict sets
    \State adopt the most conservative (maximum) $t_j^f$  
    \State solve $X_j(t)$ with boundary conditions $X_j(t_j^o), X_j(t_j^f)$ \Comment{$X_j(t)$ defined as (\ref{eq:controlFramework})}
\EndFor
\end{algorithmic}
\end{algorithm}

Obtaining the analytical solution for a constrained optimal control problem is computationally demanding and sometimes infeasible, because of the many possible combinations of the activation period of subsequent constraints \cite{Ntousakis2016}. \textcolor{r1}{
Instead of solving the otpimal control analytically as in \cite{malikopoulos2018decentralized},  Wang et al. \cite{wang2020cooperative} used a iterative process to adjust minimum safe space headway of the following vehicle to ensure collision avoidance.
}Bichiou and Rakha \cite{Bichiou2018} proposed an optimal intersection control system, which is designed to minimize travel time for CAVs. The expected distance derived from the Rakha-Parsumarthy-Adjerid (RPA) car-following model was applied as the collision avoidance constraint for each vehicle. Two versions of the proposed framework-the optimal control time and the optimal control effort-were tested on a roundabout, AWSC intersection, and a SIM.  In spite of the significant reduction in CO$_2$ emission, the proposed model has high computational expense for conducting nonlinear optimization: it takes up to five minutes to solve the optimization for a set of four vehicles. They concluded that the computational cost of solving for the optimum solution makes it impractical for real-time implementation.

\subsection{Model Predictive Control}
The model predictive control (MPC) has the advantages of dealing with a constrained system. In MPC, the optimal control problem is solved in each time step over a finite time horizon, but only control for the current time step is implemented \cite{du2018hierarchical}. Ntousakis et al. \cite{Ntousakis2016} integrated MPC control into a finite-horizon optimal control problem, where the possible real-time disturbance was compensated.  In the VICS framework proposed in \cite{kamal2015vehicle}, a risk indicator of two conflict movements was integrated into the MPC controller, along with speed error and control input.  Different from most of the studies,  the actual trajectory coordination among vehicles was implemented in the MPC framework. But the nonlinear nature of the MPC framework and its complexity does not guarantee a global optimum solution. Additionally, a good guess of the initial condition is necessary to ensure fast computation for MPC.

The optimal control framework, proposed in \cite{Bichiou2018}, had intense computational demand. As a subsequent enhancement, Bichiou and Rakha \cite{Bichiou2018b} simplified the framework by introducing MPC into the system and solved the problem numerically, instead of analytically. The improved algorithm can provide real-time solutions through convex optimization once the estimated time of arrival to the intersection is obtained. The trade-off in the slightly-reduced precision due to the convexification was justified as the authors concluded.
Du et al. \cite{du2018hierarchical} formulated the corridor-level AIM where vehicle collision was avoided by enforcing a road segment-based reference speed that is calculated by using a consensus algorithm in a decentralized manner. The control for each vehicle was formulated as a tracking system, which aimed to minimize the error between the vehicle speed and the reference speed of a particular road segment. The fast MPC method was implemented in  \cite{homchaudhuri2017fast} for the tracking system.

\subsection{Other Vehicle Control}

Lam and Katupitiya \cite{lam2013cooperative} adopted a proportional-derivative (PD) vehicle controller which was designed to maintain the gap of consecutive vehicles. The crossing sequence of vehicles was determined by a wining contest that was based on the efficiency of the crossing plan submitted by each vehicle. A lane-free AIM was proposed by Li et al. \cite{Li2018near}. Such framework is a re-imagination of intersection control by relaxing the pass-through paths of vehicles. This control problem aims to minimize the total clearance time and maximize the terminal positions at each designating leg of the intersection.  Due to the intensive computational requirement, the motion planning was decomposed into two stages. The first stage provides solutions by solving the feasible trajectory in advance and the second stage directs vehicles to form the standard formation (with equidistant row and columns) for executing the offline solution obtained in stage 1.

\section{Evaluation of AIM}
\label{sec: evalAIM}

Compared to other CAV applications, the evaluation of AIM is mostly conducted via computer simulation. In this section,  we focus on the simulation scale, benchmark comparison, measures of effectiveness, and AIM performance.

\subsection{Simulation Scale}
Recall the coverage AIM layers shown in Table \ref{table:aimLayerCoverage}. AIM approached from the traffic engineering domain typically focuses on the corridor coordination and trajectory planning layers. Carlino et al. \cite{carlino2013auction} evaluated the auction-based AIM in four major US cities with 30,000 drivers. This study was of transportation planning nature, and no coordination between intersections was made. Other studies covered mesoscopic or microscopic level of traffic, and they can be further divided into two subgroups. 

The first subgroup focused on system-optimal or near-optimal trajectory planning. No explicit vehicle dynamics model was employed.  The trajectory planning was typically formulated as a non-linear programming problem with various objectives, for example, minimum overall delay, minimum risk, maximum throughput. Their evaluation scenarios dealt with full intersection movement and with traffic demand that is close to the real world with thousands of vehicles per hour. The second subgroup approached the isolated AIM from the control engineering standpoint. Its focus is realistic vehicle dynamics model (with second- or higher-order), and the separation of conflict movements used basic requirements (e.g., FCFS policy). Additionally, the simulation scenarios are with much less traffic demand (as few as seven vehicles) and fewer movements (as few as two through movements). 

\begin{figure} [h]
	\centering
	\includegraphics[width=\columnwidth]{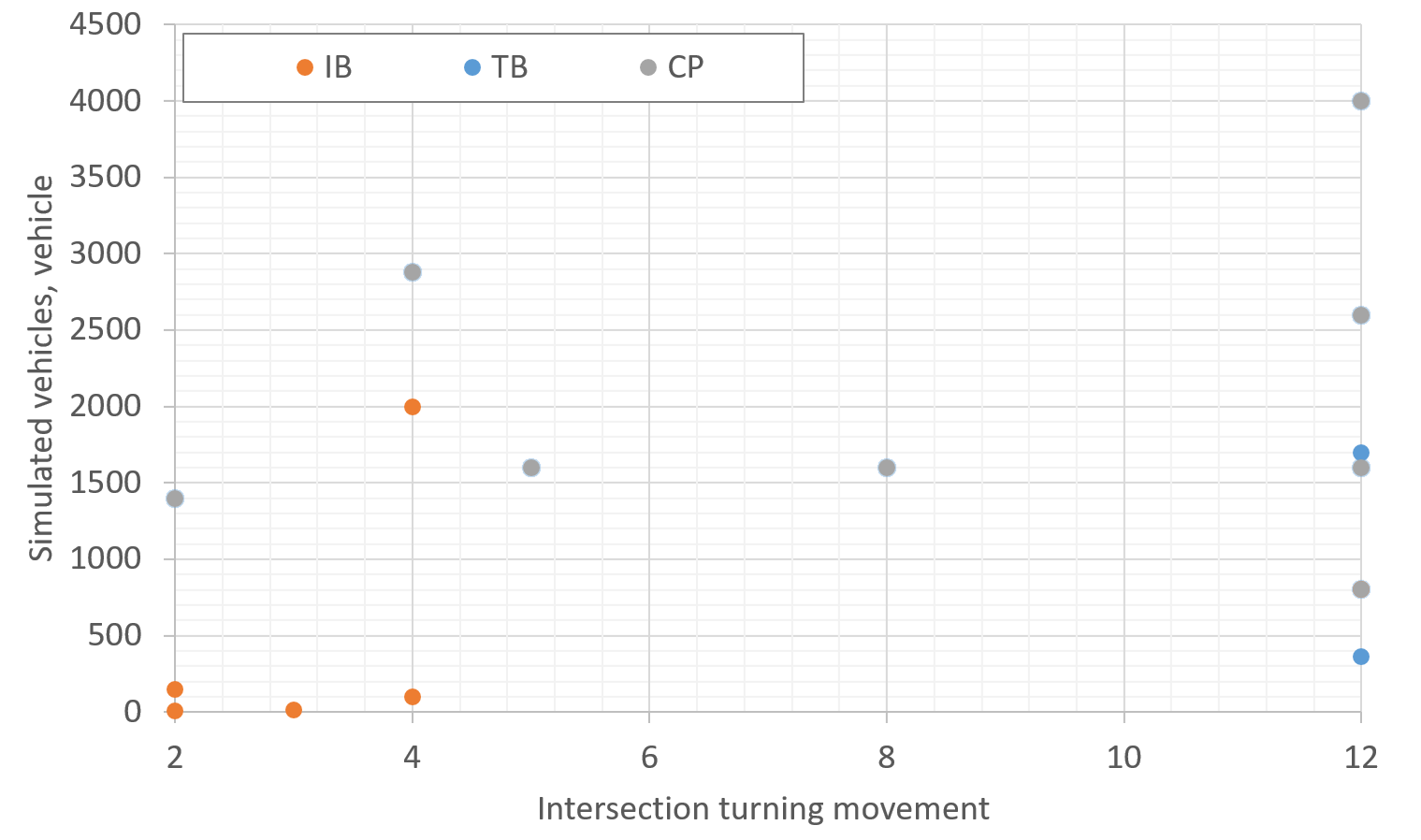}   
	\caption{Simulation scale}
	\label{fig:simVehMove}
\end{figure}

Nonetheless, an increasing amount of research from the vehicle control domain extend their interest in replacing FCFS with other forms of trajectory planning.
Fig. \ref{fig:simVehMove} displays the simulated vehicles and the number of turning movements for an intersection among the reviewed studies. The vehicle-based (VB) AIM \cite{Li2018near, Li2018a} is not shown, as it theoretically has an infinite number of turning movements. 
Due to the nature of the intersection-based (IB) reservation (occupancy of an entire intersection), the simulated movements did not excess 4 in the previous studies. The highest number of vehicles (2,000 vehicles) to be simulated for IB-AIM was conducted in \cite{Bashiri2017}. The conflict point (CP)-based and tile-based (TB) reservation were often evaluated with full (12) movements and with greater demands (upto 4,000 vehicles). 

\subsection{Benchmark Intersection Management}
\begin{figure} [h]
	\centering
	\includegraphics[width=0.8\columnwidth]{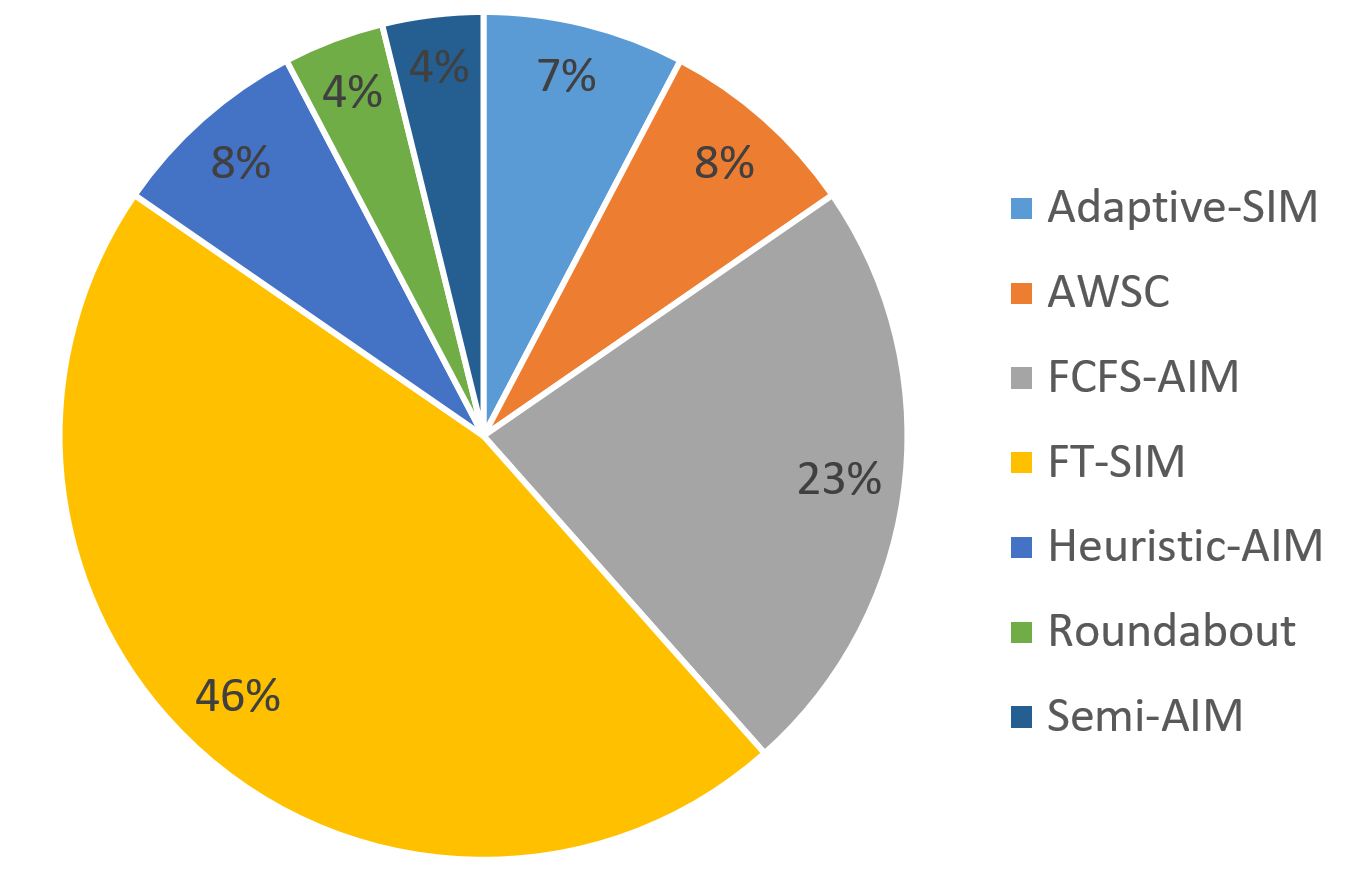}   
	\caption{Comparing intersection managements}
	\label{fig:compControl}
\end{figure}

The composition of the comparing intersection management in the reviewed studies is shown in Fig. \ref{fig:compControl}. Fixed-time SIM (FT-SIM) is the mostly-used (46\%) baseline for demonstrating the performance of proposed AIMs, followed by FCFS-AIM (23\%) and AWSC (8\%). 

Insofar,  the FT-SIMs used in the comparisons were un-optimized in accordance with traffic patterns, which may potentially limit the performance of FT-SIM. As shown with decades of practices in traffic engineering \cite{koonce2008traffic, manual2016highway, urbanik2015signal}, an optimized SPaT plan based on prevailing traffic patterns could significantly improve intersection performance. The optimization relies on the balance between cycle time, phase time, queue, etc. Furthermore, a better comparison for SIM can be found with actuated SIM \cite{mirheli2018development} or adaptive SIM \cite{liu2019trajectory}.  
The second most-frequent intersection management used for comparison was FCFS-AIM, which accounted for 23\% in the previous studies. FCFS is the default priority rule of AIM, hence FCFS-AIM was often adopted as benchmark for evaluating subsequent AIM variants: auction-based  \cite{carlino2013auction}, control input-based (acceleration) \cite{creemers2018design}, platoon-based \cite{jin2013platoon} AIMs, etc.
AWSC is the third most-used baseline, largely due to its similar nature with AIM: 1) the priority of crossing is assigned at a vehicle level; 2) it is unsignalized; and 3) it operates on a FCFS basis. However, AWSC is a relatively inefficient scheme, as the required stop for each vehicle can potentially increase the delay and queue. At median demand, SIM is recommended as per the Manual on Uniform Traffic Control Devices \cite{MUTCD2009}.

\subsection{Measure of Effectiveness}

The most common measures of effectiveness (MOEs) for assessing intersection performance are average delay, average fuel consumption, average CO2 emission, travel time, evacuation time, intersection throughput, as shown in Table \ref{table:moeTable}.

\begin{table*}[h]
\centering
\caption{Comparison of AIM Studies}
\begin{tabular}{lcccccccc}
\hline
\multirow{2}{*}{Study} & \multicolumn{3}{c}{Mobility} & \multicolumn{2}{l}{Environmental} & \multicolumn{1}{c}{\multirow{2}{*}{Comfort}} & \multicolumn{1}{c}{\multirow{2}{*}{\begin{tabular}[c]{@{}c@{}}Computational \\ Framework\end{tabular}}} & \multicolumn{1}{c}{\multirow{2}{*}{Other}} \\ \cline{2-6}
 & \multicolumn{1}{c}{Delay} & \multicolumn{1}{c}{Evac Time} & \multicolumn{1}{c}{Throughput} & \multicolumn{1}{c}{CO2} & \multicolumn{1}{c}{Fuel} & \multicolumn{1}{c}{} & \multicolumn{1}{c}{} & \multicolumn{1}{c}{} \\ \hline
Bichiou and Rakha \cite{Bichiou2018b}  & \checkmark &  &  &  \checkmark & \checkmark &  &  &  \\ \hline
Liu et al. \cite{liu2019trajectory} &  \checkmark &   \checkmark &  \checkmark &  &  &  &  &  \\ \hline
Ding et al.\cite{Ding2017a} &  &   \checkmark &   \checkmark &  \checkmark &  \checkmark &   \checkmark&  &  \\ \hline
Fajardo et al. \cite{Fajardo2011}  &  \checkmark &  &  &  &  &  &  &  \\ \hline
Hassan and Rakha \cite{hassan2014fully}  &  \checkmark &  &  &  &  &  &  &  \\ \hline
Kamal et al. \cite{kamal2015vehicle}   &  &  \checkmark &  &   & \checkmark  &  &  &  \\ \hline
Stone et al. \cite{stone2015autonomous}  & \checkmark  &  &  &  &  &  &  &  \\ \hline
Levin and Rey \cite{levin2017conflict} &  & \checkmark  &  &  &  &  &  &  \\ \hline
Lam and Katupitiya \cite{lam2013cooperative}  & \checkmark  &  &  &  &  &  &  &  \\ \hline
Li et al. \cite{Li2018near}  &  &  &  &  &  &  &  \checkmark   &  \\ \hline
Du et al. \cite{du2018hierarchical} &  &  \checkmark  &  &  &  &  &  &  \\ \hline
Jin et al. \cite{jin2013platoon}  &  &  \checkmark  &  &  &  \checkmark   &  &  &  \checkmark  \\ \hline
Bashiri and Fleming \cite{Bashiri2017}  &  \checkmark  &  &  &   &   &  &  &  \\ \hline
Bashiri et al. \cite{bashiri2018paim}  &   \checkmark &  &  \checkmark  &  &  \checkmark  &  &  &  \\ \hline
M{\"u}ller et al. \cite{muller2016intersection}  &  &  &  \checkmark  &  &  &  &  &  \\ \hline
Mirheli et al. \cite{mirheli2018development}  &  \checkmark   &   \checkmark  &  &  &  &  &  &   \\ \hline
Carlino et al. \cite{carlino2013auction} &  & \checkmark  &  &  &  &  &  &   \\ \hline
Creemers et al. \cite{creemers2018design}  &  &  &  &  &  &  &  &  \checkmark  \\ \hline
Jiang et al. \cite{jiang2017distributed}  &  &  &  &  &  &  &  \checkmark  &  \\ \hline
\end{tabular}
\label{table:moeTable}
\end{table*}

Among them, delay is the measure that relates to drivers' experience the most, as it represents the excessive amount of time in traversing an intersection.  Delay can be further broken down to stopped time delay, approach delay, travel time delay, time-in-queue delay, and control delay \cite{Mathew2014}.  Though analytical delay prediction models (e.g., Webster's, Akcelik, HCM2000) have been proposed along the years, simulation provides an innovative and robust way of evaluating the delays for intersections. 
Queue length provides an indication of whether a given intersection impedes the vehicle discharging from an upstream intersection. Queue length is typically taken into account for SIM coordination. For AIM, since vehicle crossing is scheduled at an individual level, the queue length becomes less effective in representing the overall intersection performance. 

The number of stops is an important parameter when it comes to emission model, since the regaining of speed form a stopped vehicle requires additional acceleration, therefore burning more fuel. The use of fuel consumption model has become more prevailing, such as the VT-Micro model \cite{rakha2004development}. Other less common emissions have been adopted for studies as well. Hydrocarbon (HC), carbon monoxide (CO), nitric oxide (NO), and fuel consumption were used by \cite{chou2014evaluation} for evaluating the proposed triangabout. The average speed and ratio of averaging moving time are used by \cite{hildebrand2007unconventional} and \cite{reid2001travel}, respectively. Safety surrogate assessment measure (SSAM) is often used to gauge the safety performance for human-driven vehicles. However, its applicability to CAVs is still open to debate, because the performance of CAV is expected to exceeed the physiological limitations of human drivers (e.g., much quicker response time). Evaluation time for a fixed amount of vehicles and the minimal entry time of consecutive vehicles in the priority queue were also used by \cite{Li2018near} and \cite{malikopoulos2018decentralized}, respectively. 
Other innovative comparing metrics had also been adapted.  The required number of iterations in the proposed distributed AIM was used as a performance measure in \cite{jiang2017distributed}. The time required for stabilizing the intersection queue length to a minimal level was used as the MOE in \cite{creemers2018design}.

\subsection{AIM Performance}

\begin{figure}[h]
\begin{minipage}[h]{1\columnwidth}
\centering
\subfloat[Metric to Decrease]{\includegraphics[scale=0.135]{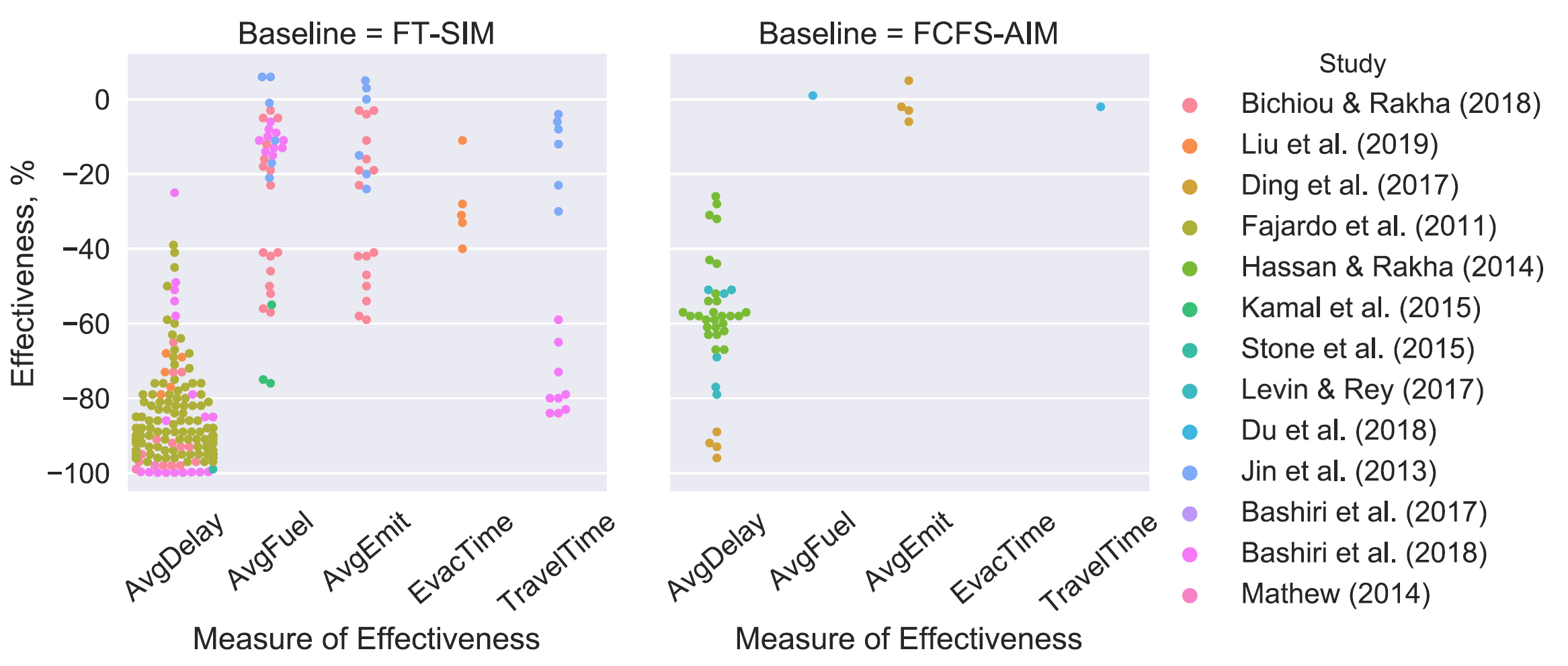}}
\end{minipage} 
\begin{minipage}[h]{1\columnwidth}
\centering
\subfloat[Metric to Increase]{\includegraphics[scale=.13]{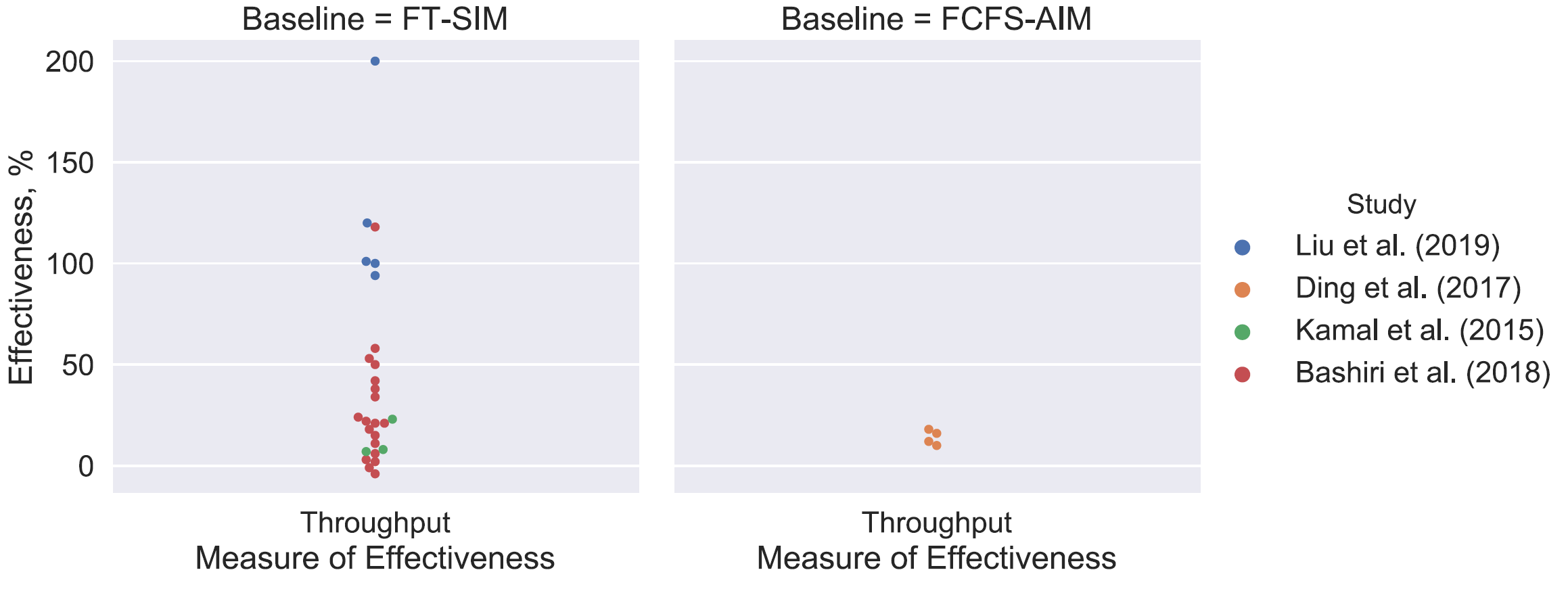}}
\end{minipage} \par
\caption{Measures of effectiveness}
\label{fig:moe}
\end{figure}

Fig. \ref {fig:moe} offers an overview of the performance gain in the six most common MOEs. Each sample point represents a scenario-based comparison. As shown, some studies conducted multiple comparisons, while others only did fewer. The left-hand-side figure shows the comparison of proposed AIMs and FT-SIMs; whereas the one on the right-hand side exhibits the comparison between proposed AIMs and the FCFS-AIMs, the default configuration for AIM. Fig. \ref {fig:moe}(a) plots the MOEs that one aims to decrease (e.g., delay, fuel consumption), and Fig. \ref {fig:moe}(b) show the MOE that one tries to increase, such as intersection throughput.
For the papers without explicit numeration of the MOEs for all the scenarios (e.g., only figures was shown), we employed a plot trace tool \cite{rohatgi} to extract the numeric values in the original figures.

Two patterns can be observed in Fig. \ref {fig:moe}. First, the average delay has been adapted more frequently than any other MOE. Second, the percentage of gain or loss spans a wide range, for instance, from 6\% (increase) down to nearly -100\% (decrease). The throughput of the proposed AIM in \cite{liu2019trajectory} was doubled (increased to 200\%). Based on the reported results, the AIM performs exceptionally well.

Besides the benefits attributed to AIM, the evaluation scenario (e.g., turning movements, traffic demand) likely plays a significant role in obtaining such high performance.
Nevertheless, there are several factors that could potentially skew performance. The first factor is the traffic pattern, mainly the level of saturation of the intersection. The saturation flow rate of a signalized intersection lies within the range of 1,700-1,900 vph with variations \cite{manual2016highway}. Due to the safety separation among crossing vehicles, it is not hard to imagine a scenario where a vehicle has to slow down, even to a complete stop at high traffic density to maintain safe separation. It is reasonable to believe that the performance of an AIM could be impacted under such circumstances.
Second, the benchmark intersection management scheme could play a significant role in comparison. At median or high traffic demand, an optimized FT-SIM can substantially outperform its non-optimized counterpart. 
Hence, the SIM should be calibrated or optimized to increase the validity of the comparison.
Additional realistic traffic scenarios for AIM are needed in order to draw a statistically sound conclusion for the applicability between AIM and conventional SIM.


\section{Research Trends and Topics}
\label{sec: discussion}
%
Intersections are the common traffic bottlenecks in the modern transportation network. In this section, we discuss future research directions for AIM.

\subsubsection{Benchmarking}

The majority of the studies served as proof-of-concept studies with two primary focuses: safe operation and potential benefits.
Based on the reported results, the AIM performs exceptionally well (i.e., nearly 100\% reduction in average delay in certain cases) under low demand scenario without the consideration of lane change activity. Thus far, there is no consensus on a benchmark scenario for reliably and consistently assessing the performance of AIM. For SIM, the Highway Capacity Manual \cite{manual2016highway} provides a detailed methodology for the optimization of SPaT plans according to the local traffic pattern. For AIM, however, only few proposed ones were subject to a wide variety of scenarios. Investigations regarding the carrying capacity of AIMs are very much desired.

\subsubsection{Semi-autonomous Intersection Management}
The Volpe National Transportation Systems Center \cite{volpe2008vehicle} estimated that it might take 25-30 years for CAVs to reach a 95\% market penetration rate (MPR), even with federally mandatory installation of DSRC devices on new light vehicles manufactured in the United States. In anticipation of the transition to full penetration, semi-AIM which integrates non-equipped vehicles is a logical step. So far, there are a few Semi-AIMs that have been proposed. The potential compromise needs to carefully balance between compatibility and performance. The semi-AIM with signalization proposed by Dresner and Stone \cite{dresner2007sharing} suffered from significant degradation in performance even with only 5-10\% presence of human operation. It was found that the CAVs were blocked by human-driven vehicles that are controlled by the signalization. An enhanced version was proposed in \cite{stone2015autonomous} to relax this limitation, where the control of a vehicle can be provisionally transferred to the ADAS. Furthermore, the traffic signal also acted as a fallback strategy when a reservation cannot be obtained. Under the semi-AIM framework, similar performance as AIM was achieved with no more than 40\% MPR. However, additional research is still required to quantify the possible trade-off that has to make when it comes to semi-AIM.

\subsubsection{Priority Policy}
AIM enables priority to be assigned at individual vehicle level, compared to SIM that operates at a vehicle group movement level. Insofar, the majority of the proposed AIMs adopted FCFS policy, which is not likely to lead to system optimum if solely implemented. Utilizing other network level information (e.g., routing) seems an good complement to the FCFS policy for its further enhancement \cite{levin2016paradoxes}. Besides FCFS, other non-fairness-based policies have shown their potentials for AIM, which could be based on system optimality, travel mode, game theory, or custom priority score system. Nevertheless, the priority policy remains an underexplored area, and it is crucial in increasing applicability of AIM to a wide variety of traffic operation goals. 

\subsubsection{Computational Efficiency}
Among the four types of reservation systems for AIM, the vehicle-based, free-movement reservation holds the greatest amount of vehicles in theory. However, its high computational complexity hinders it from being implemented in real-time. The practicality of the game-theoretic reservation remains an open question according to the literature. The tile-based and the conflict point-based reservations are the better candidates for real-world implementation, though their computation is still nontrivial. Intense computation is still one of the major hurdles for AIM. Hence, the computational efficiency of the reservation system desires much enhancement. The decomposition of intersection management and vehicle control in several studies has exhibited great potentials in improving computational efficiency. 

\subsubsection{Applicability with Intersection Layout}
A generalized reservation-based algorithm is needed as not all the intersections in the real world follow the standard symmetrical 4-leg layout. Such variations may require necessary modifications of the existing framework. Besides roundabout, there is an increasing amount of alternative intersections that have been implemented in the real world \cite{zhong2018unconventional}. For example, the diverging diamond interchange (DDI) has an unconventional layout where the through movements on the major directions intersect twice (resulting in two conflict points), instead of once. Intersection-based reservation is rather counterproductive in the case of DDI, as the interchange could span 300-meter long. Therefore, the suitability of reservation systems to the intersection layout is worthy of investigation. In the DDI case, the conflict point-based reservation systems are likely to function better than the tile-based reservation system.

\subsubsection{Cyber-security of CAV}
More emphasises have been put on the cyber-security aspect of CAV. As a communication platform, CAV is susceptible to both passive and active forms of malicious attack.  Passive attack has lower risk, for instance, the eavesdrop of information of a target vehicle \cite{elliott2019recent}. Active attack may include spoofing incorrect data \cite{chen2018exposing}, unauthorized message modification, denying of service, and GPS jamming. A recent study conducted by Chen et al. \cite{chen2018exposing} illustrated that the computational complexity should be factored in when it comes to algorithm design for intersection management. Even though the I-SIG system \cite{ahn2016multimodal} in the study is robust in theory, it is subject to exploitation in practice when only a simplified version of the algorithm can be used due to computational limitation of the RSU. Therefore, security research should also extend to the design of an AIM algorithm.

\subsubsection{Decentralization}
Centralized control is often subject to single-point failures, making it a worthwhile target for attackers. The current information computing infrastructure exhibits the trend of decentralization, and we have seen an increasing amount of decentralized vehicle control for AIMs. With the increased amount of computational power, CAV is capable of computing and analyzed location traffic information with the onboard unit before sending actionable information to AIM. This trend on vehicle control coincides with the concept of edge computing. Other latest technologies can evolve AIM to evolve into a decentralized, robust system further. For instance, blockchain technology enables the tamper resistance for any transaction that is stored by each CAV \cite{leiding2016self, li2019blockchain}.

\section{Conclusion}
\label{sec:conclusion}
Managing traffic safely and efficiently at intersections remains one of the most challenging problems for our transportation system. The CAV technology extends the intersection management down to individual vehicle control, offering a new degree of flexibility to meet operational goals. This paper systematically reviews the state of the research of autonomous intersection management (AIM). The intersection management is distilled into three hierarchical layers, which are corridor coordination layer, intersection management layer, and vehicle control layer. The underlying design concepts for AIM are discussed in details. Additionally, the necessary connection to existing signalized intersection management is also made to present a full picture of the overall intersection management. This review shows that the reservation system with high computational efficiency and the extension to vehicle control level are two active, and yet underexplored, areas. Also, a consensus on the evaluation scenario for AIMs is necessary to accelerate the AIM research. Lastly, this paper highlights key future research topics from an interdisciplinary standpoint. 

\appendix
\setcounter{figure}{0}
\setcounter{table}{0}   
\renewcommand{\thetable}{A\arabic{table}}
\label{appendix}
 
\subsection{\textcolor{r1}{List of Abbreviations}}
\begin{table}[H]
\centering
\caption{\textcolor{r1}{List of Abbreviations}} 
\begin{tabular}{p{0.6in}|p{2.2in}} 
\hline  \hline
\textbf{Abbreviation} & \textbf{Definition} \\ \hline 
ADAS & advanced driver-assistance systems \\ \hline
ADS & automated driving systems \\ \hline
AIM & autonomous intersection management \\ \hline 
AV & automated vehicles  \\ \hline 
AWSC & all-way stop control  \\  \hline  
CV & connected vehicles \\ \hline
CAV & connected and automated vehicles\\ \hline
CACC-CG & cooperative adaptive cruise control-cooperative game \\ \hline
CVIC & cooperative vehicle intersection control \\ \hline
CP & conflict point-based \\ \hline
DDI & diverging diamond interchange \\ \hline
DSRC & dedicated short-range communication \\ \hline
FIFO & first-in-first-out \\ \hline    
FCFS & first-come-first-serve \\ \hline
FT-SIM & fix-time signalized intersection management \\ \hline
HCM & Highway Capacity Manual \\ \hline
HV & human-driven vehicle  \\  \hline  
IB & intersection-based \\ \hline
I-SIG & Intelligent Traffic Signal System \\ \hline
MILP & mixed-integer programming \\ \hline
MPC & model predictive control \\ \hline
MPR & market penetration rate\\ \hline
MOE & measure of effectiveness \\ \hline
NLP & non-linear programming \\ \hline
RSU & roadside unit \\ \hline
SICL & signal-free intersection control logic \\ \hline
SPaT & signal phase and timing    \\ \hline
SIM & signalized intersection management \\ \hline
SAE & Society of Automotive Engineers International \\ \hline  
TB & tile-based \\ \hline
VB & vehicle-based \\ \hline
VICS & vehicle-intersection coordination scheme \\ \hline
V2I & vehicle-to-infrastructure \\ \hline
V2X & vehicle-to-everything \\ \hline
\end{tabular}
\label{table:abbrv}
\end{table}

\bibliographystyle{IEEEtran}
\bibliography{FINAL_VERSION}

\newpage
\begin{IEEEbiography}
    [{\includegraphics[width=1in,height=1.25in,clip,keepaspectratio]{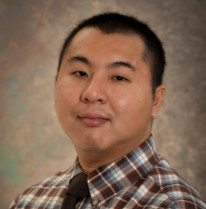}}]{Zijia (Gary) Zhong} is a Postdoctoral Researcher in the Center of Integrated Mobility Sciences at the National Renewable Energy Laboratory, United States. He received his Ph.D degree in Transportation Engineering and Master's degrees in Civil Engineering from the New Jersey Institute of Technology in 2018 and 2011 respectively. His research interests include deployment of intelligent transportation systems (ITS), emerging mobility, high-performance computing for transportation modeling, data analytics, vehicle platooning, highway automation, and human factor study for ADAS.
\end{IEEEbiography}

\begin{IEEEbiography}
    [{\includegraphics[width=1in,height=1.25in,clip,keepaspectratio]{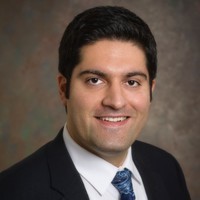}}]{Mark Nejad}
is an Assistant Professor in the Department of Civil and Environmental Engineering at the University of Delaware. His research interests include connected and automated vehicles, network optimization and control, and game theory. He has published more than thirty peer-reviewed papers in venues such as Transportation Science, IEEE Transactions on Parallel and Distributed Systems, and IEEE Transactions on Computers. He received several publication awards including the 2016 Institute of Industrial and Systems Engineers (IISE) Pritzker Best Doctoral Dissertation Award and the INFORMS ENRE best student paper award. He is a member of the IEEE and INFORMS. 
\end{IEEEbiography}

\begin{IEEEbiography}
    [{\includegraphics[width=1in,height=1.25in,clip,keepaspectratio]{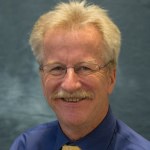}}]{Earl (Rusty) E. Lee, II}
received his Bachelor’s degree in Nuclear Engineering, Master’s degree in Management, and Ph.D. degree in Decision Sciences and Engineering Systems from Rensselaer Polytechnic Institute in 1978, 2004, and 2006, respectively.  Currently, he is an Assistant Professor in the Department of Civil and Environmental Engineering at the University of Delaware. He is also the Director of the Delaware Technology Transfer (T2) Center and a Core Faculty member in the Disaster Research Center. His research interests include disaster management, infrastructure system modeling, and transportation system operations.
\end{IEEEbiography}

\end{document}